\documentclass[pra,twocolumn,showpacs]{revtex4}
\usepackage{epsfig}
\usepackage{amsmath,supertabular}
\usepackage{amssymb,mathrsfs}
\usepackage[debug=true,hypertex,linkcolor=red]{hyperref}
\newtheorem{Lem}{Lemma} \newtheorem{Cor}{Corollary}
\newtheorem{Theo}{Theorem}
\newtheorem{Def}{Definition}
\newtheorem{Prop}{Proposition} \def\Supp{\mathsf{Supp}}
\def\map#1{\mathscr #1}

 \def\qed{$\blacksquare$ \medskip}
\def\Proof{{\bf Proof. }}  \def\Lin#1{\mathsf{Lin}\left( #1\right)}
\def\Tr{\mathrm{Tr}} \def\>{\rangle} \def\<{\langle}
\def\kk{\rangle\!\rangle}\def\bb{\langle\!\langle}
\def\spc#1{\mathcal{#1}} 

\def\newColor #1 {\expandafter\def\csname text#1\endcsname{}\expandafter\def\csname #1\endcsname##1{\special {color
      push #1}##1}}%
 \newColor
Aquamarine 
\begin{document}
\title{A short impossibility proof of Quantum Bit Commitment}
\author{Giulio Chiribella} \email{chiribella@fisicavolta.unipv.it}
\affiliation{Quantum Information Theory Group, Dipartimento di Fisica
  A.~Volta, via Bassi 6, 27100 Pavia, Italy} 
\author{Giacomo Mauro
  D'Ariano} \email{dariano@unipv.it} \affiliation{Quantum Information
  Theory Group, Dipartimento di Fisica A.~Volta, via Bassi 6, 27100
  Pavia, Italy} 
\author{Paolo Perinotti}
\email{perinotti@fisicavolta.unipv.it} \affiliation{Quantum
  Information Theory Group, Dipartimento di Fisica A.~Volta, via Bassi
  6, 27100 Pavia, Italy} 
\author{Dirk Schlingemann}
\email{d.schlingemann@tu-bs.de} \affiliation{ISI Foundation, Quantum
  Information Theory Unit, Viale S. Severo 65, 10133 Torino, Italy}
\author{Reinhard F.~Werner} \email{Reinhard.Werner@itp.uni-hannover.de}
\affiliation{Institut f\"ur Theoretische Physik, Leibniz
  Universit\"at Hannover, Appelstrasse 2, 30167 Hannover, Germany} 
\date{\today}

\begin{abstract}
  Bit commitment protocols, whose security is based on the laws of
  quantum mechanics alone, are generally held to be impossible on the
  basis of a concealment-bindingness tradeoff \cite{LC97,May97}. A
  strengthened and explicit impossibility proof has been given in Ref.
  \cite{werqbc} in the Heisenberg picture and in a C${}^*$-algebraic
  framework, considering all conceivable protocols in which both
  classical and quantum information are exchanged. In the present paper
  we provide a new impossibility proof in the Schr\"{o}dinger picture,
  greatly simplifying the classification of protocols and strategies
  using the mathematical formulation in terms of quantum combs
  \cite{qca}, with each single-party strategy represented by a {\em
    conditional comb}. We prove that assuming a stronger notion of
  concealment---worst-case over the classical information
  histories---allows  Alice's cheat to pass also the
  worst-case Bob's test. The present approach allows us to restate the
  concealment-bindingness tradeoff in terms of the continuity of
  dilations of probabilistic quantum combs with respect to
  the comb-discriminability distance.
\end{abstract}
\maketitle

\section{Introduction}\label{sec:intro}

Bit commitment involves two mistrustful parties---Alice and Bob---of
which Alice submits to Bob a piece of evidence that he will use to
confirm a bit value that she will later reveal, whereas Bob cannot
determine the bit value from the evidence alone. A good bit commitment
protocol should be simultaneously {\em concealing} and {\em binding},
namely the evidence should be submitted to Bob in such a way that he
has (almost) no chance to identify the committed bit value before
Alice later decodes it for him, whereas Alice has (almost) no way of
changing the value of the committed bit once she has submitted the
evidence. In the easiest example to illustrate bit commitment, Alice
writes the bit down on a piece of paper, which is then locked in a
safe and sent to Bob, whereas Alice keeps the key. At a later time,
she will unveil the bit by handing over the key to Bob. However, Bob
may be able to open the safe in the meantime, and this scheme is in
principle insecure. Yet all bit commitment schemes currently used rely
on strongboxes and keys made of computations that are (supposedly)
hard to perform (see Ref. \cite{werqbc} for a list of references), and
cryptographers have long known that bit commitment (like any other
interesting two-party cryptographic primitive) cannot be securely
implemented with classical information \cite{Kil88}.

Besides having immediate practical applications, bit commitment is
also a very powerful cryptographic primitive. Conceived by Blum
\cite{Blu83} as a building block for secure coin tossing, it also
allows to implement secure oblivious transfer
\cite{BBC+91,Cre94,Yao95}, which, in turn, is sufficient to establish
secure two-party computation \cite{Kil88,CVT95}.

It has therefore been a long-time challenge for quantum cryptographers
to find {\em unconditionally secure} quantum bit commitment protocols,
in which---very much in parallel to quantum key distribution
\cite{BB84,Eke91}---security is guaranteed by the laws of quantum
physics alone.

The first quantum bit commitment protocol appeared in the famous Bennett and Brassard 1984 quantum
cryptography paper \cite{BB84}, in a version for implementing coin tossing. However, they also
proved that Alice can cheat using EPR correlations, by which she can unveil either bit at the
opening stage by measuring in the appropriate basis a particle entangled with the one encoding the
bit, whereas Bob has no way to detect the attack. Subsequent proposals for bit commitment schemes
tried to evade this type of attack, {\em e.g.} in the protocol of Ref. \cite{BCJ+93} which for a
while was generally accepted to be unconditionally secure.

In 1996 Lo and Chau \cite{LC97}, and Mayers \cite{May97} realized that all previously proposed bit
commitment protocols were vulnerable to a generalized version of the EPR attack that renders the
BB84 proposal insecure, a result that they slightly extended to cover quantum bit commitment
protocols in general. Their basic argument is the following. At the end of the commitment phase, Bob
will hold one out of two quantum states $\varrho_k$ as proof of Alice's commitment to the bit value
$k \in \{0,1\}$.  Alice holds its purification $\psi_k$, which she will later pass on to Bob to
unveil. For the protocol to be concealing, the two states $\varrho_k$ should be (almost)
indistinguishable, $\varrho_0 \approx \varrho_1$. But Uhlmann's theorem \cite{Uhl76} then implies
the existence of a unitary transformation $U$ that (nearly) rotates the purification of $\varrho_0$
into the purification of $\varrho_1$. Since $U$ is localized on the purifying system only, which is
entirely under Alice's control, Lo-Chau-Mayers argue that Alice can switch at will between the two
states, and is not in any way bound to her commitment.  As a consequence, any concealing bit commitment
protocol  is argued to be necessarily non-binding (these results still hold true
when both parties are restricted by superselection rules \cite{KMP04}). So while the proposed
quantum bit commitment protocols offer good practical security on the grounds that Alice's EPR
attack is hard to perform with current technology, none of them is unconditionally secure.

Starting from 2000 the Lo-Chau-Mayers no-go theorem \cite{LC97,May97} has been continually
challenged by Yuen and others \cite{Yuenall,Yuenlast,Che01}, arguing that the impossibility proof of Ref.
\cite{LC97} does not exhaust all conceivable quantum bit commitment protocols, whereas it is still
unclear if Mayer's framework \cite{May97} is complete. Several protocols have been proposed and
claimed to circumvent the no-go theorem \cite{Yuenall}. These protocols seek to strengthen Bob's
position with the help of `secret parameters' or `anonymous states', so that Alice lacks some
information needed to cheat successfully: while Uhlmann's theorem would still imply the existence of
a unitary cheating transformation as described above, this transformation might be unknown to Alice.

The above attempts to build up a secure quantum bit commitment protocol have motivated the thorough
analysis of Ref. \cite{werqbc}, which provided a strengthened and explicit impossibility proof
exhausting all conceivable protocols in which classical and quantum information is exchanged between
two parties, including the possibility of protocol aborts and resets. The proof \cite{werqbc}
encompasses protocols even with unbounded number of communication rounds (it is only required that
the expected number of rounds is finite), and with quantum systems on infinite-dimensional Hilbert
spaces.  However, the considerable length of the proof in Ref. \cite{werqbc} makes it still hard to
follow (see {\em e.g.}  comments in Ref.  \cite{Yuenlast}), lacking a synthetic intuition of the impossibility proof.

The debate can be only settled with an appropriate formulation of the
problem, which is sufficiently powerful to include all possible
protocols in a single simple mathematical object, thus leaving no
shadow of doubt on the completeness of the protocol classification.
Once the mathematical formulation of all protocols is settled, then
the impossibility statement becomes just a mathematical theorem. In
this paper we will first see that the appropriate notion to describe
all individual strategies in a purely quantum protocol is the {\em
  quantum comb}. The quantum comb generalizes the notion of quantum
operation of Kraus \cite{Kraus}, and has been originally introduced in
Ref. \cite{qca} to describe quantum {\em circuit boards}, where inputs
and outputs are not just quantum states, but quantum operations
themselves. Since quantum combs are in one-to-one correspondence with
sequences of quantum operations \cite{qca,comblong}, a quantum comb is
suited to represent the sequence of moves performed by a party in a
multi-round quantum protocol.  Indeed, the same mathematical structure of quantum combs has been recognized by Gutoski and Watrous in
Ref.  \cite{watrous} as the appropriate formulation of multi-round
quantum games.  In order to treat protocols that involve both quantum
and classical communication, we will then extend this framework by
introducing the concept of \emph{conditional comb}, which describes a
computing network that is able to sequentially process both quantum and
classical information.

\begin{figure}[h]
  \epsfig{file=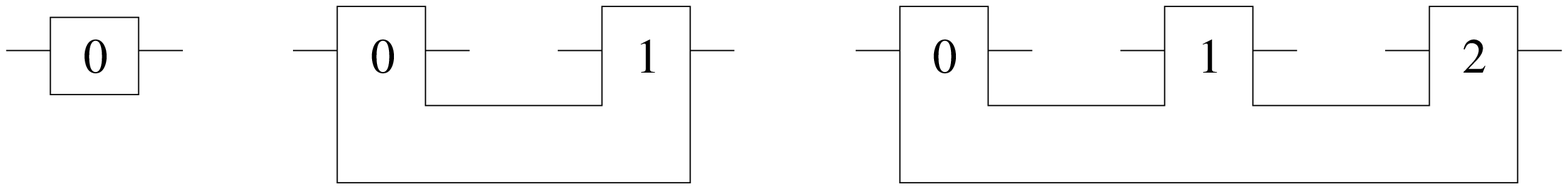,width=8cm}
  \caption{\label{f:qcombs}  Diagrammatic
    representation of quantum combs.  For a quantum comb each line
    entering or exiting a tooth represents a quantum system,
    while for a conditional comb it represents a hybrid
    quantum-classical system. A quantum operation (the box on the left
    upper corner) is a special case of quantum comb with a single
    tooth.}
\end{figure}

\begin{figure}[h]
  \epsfig{file=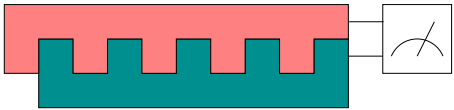,width=6cm}
  \caption{A two-party protocol in which classical and quantum
    information are exchanged assigns the set of allowed conditional
    combs, along with the pertaining input-output structure. A
    conditional comb is a collection of quantum combs labeled by
    histories of classical communication, each quantum comb
    representing a specific sequence of single-party moves for a
    particular classical history. Each tooth of a comb corresponds to
    a single turn of the protocol, the last one representing the last
    turn in the commitment phase.  For histories ending in a
    successful commitment, at the opening Bob performs a joint
    measurement on all systems available to him.  Combining Bob's
    comb before the opening with this final measurement yields a
    special case of quantum comb---the so-called {\em quantum
      tester}---whose output is the committed bit value. In this
    framework, Alice's comb plays the role of a ``state'' encoding the
    bit value, whereas Bob's tester plays the role of a ``POVM'' for
    binary discrimination.  Such binary discrimination---prescribed by
    the protocol at its end---should not be confused with Bob's
    attempts to discriminate Alice's strategies before the opening.}
  \label{f:interlacedqcombs}
\end{figure}

Examples of combs are represented diagrammatically in Fig.  \ref{f:qcombs}. For a purely quantum
comb, each line entering or exiting a tooth of the comb represents a quantum system. For a
conditional comb, each line represents a hybrid quantum-classical system, accounting also for
classical information exchanged at each step.  In a two-party protocol, a comb represents a
single-party strategy, with each tooth of the comb representing the move performed by the party at
some turn.  Subsequent turns are represented by subsequent teeth, from left to right.  The output oft
the multi-round protocol is given by two combs interlaced as in Fig.  \ref{f:interlacedqcombs}---the
upper Bob's, the lower Alice's.  The exchange of quantum-classical systems can be mathematically
described in a C${}^*$-algebraic representation of a deterministic comb, or, equivalently, by
treating the conditional comb as a collection of (purely quantum) probabilistic combs, each of them
being labeled by a particular history of classical communication. In this paper, we will choose the
second point of view, which avoids using the C${}^*$-algebraic framework, with the need, however, of
considering collections of probabilistic quantum combs, accounting for the classical information
coming from measurements.

A protocol assigns the set of allowed strategies, i.e. the set of allowed conditional combs, along
with the pertaining input-output structure regulating the exchange of quantum and classical
information.  As already mentioned, a conditional comb is a collection of probabilistic quantum
combs, each of them representing the sequence of single-party moves associated to a particular
history of classical communication.  In a general protocol, some histories will lead to a successful
commitment, while some other will possibly lead to an {\em abort}, in which the two parties
irrevocably give up, excluding any further communication (if the protocol is restarted, then the
concatenation of the two sequences can be regarded as part of a new longer protocol with possible {\em
  resets}). Accordingly, we will consider histories from the beginning to the end of the commitment
(which can be either successful or not), {\em i.e} excluding the opening.  Each tooth of a comb
corresponds to a single turn of the protocol, and, in the case of successful commitment, the last
tooth represents the last turn before the opening.

For histories that end in a successful commitment, in the opening
Alice will send to Bob a classical message along with a set of
ancillae prescribed by the protocol, and Bob will perform a suitable
joint measurement on all quantum systems available to him, as in Fig.
\ref{f:interlacedqcombs}.  The combination of Bob's comb (up to the
opening) with the final measurement at the opening is itself a special
case of quantum comb---the so-called {\em quantum tester}---whose
output is the committed bit value.  In this framework, Alice's comb
plays the role of a ``state'' encoding the bit value, whereas Bob's
tester plays the role of a ``POVM'' for binary discrimination.  Such
binary discrimination---prescribed by the protocol at its end---should
not be confused with Bob's attempts to discriminate Alice's strategies
before the opening.  We will see that the fact that the protocol has
many rounds actually can help Bob in discriminating between different
Alice's strategies. Thus, the probability of Bob cheating---which in a
protocol with a single Bob-Alice-Bob round would be represented by the
CB-norm distance between Alice's channels---here is replaced by the
{\em comb distance}\cite{memorydisc}, which is typically larger than the
CB-norm, since Bob can exploit the memory structure of Alice's
strategy.

In the following we will consider the concealment-bindingness tradeoff
for any possible history of classical information exchanged within the
protocol. This will allow us to restate the tradeoff in terms of a
mathematical theorem assessing the continuity of dilations of
probabilistic quantum combs in terms of the comb
discriminability-distance.  The dilation theorem states that any
probabilistic comb can be dilated to a sequence of single-Kraus
quantum operations, upon introducing some additional ancillae.  As a
consequence, the impossibility proof will run essentially as follows.
At the end of the commitment phase, two possible Alice's strategies
for the bit values 0 and 1, respectively, are (almost) indistinguishable to Bob, who lacks the quantum information encoded in Alice's ancillae.
Instead, at the opening, the two dilated strategies of Alice
corresponding to the two values of the committed bit are (almost)
perfectly discriminable. As a consequence of indstinguishability up to
the opening phase, Alice can choose between the two strategies by
performing a unitary transformation on the ancilla in the last tooth
of her comb.  Therefore, one has (almost) perfect opening, and, at the
same time, Alice can cheat perfectly. The concealment-bindingness
tradeoff is thus reduced to the continuity of the dilation of
probabilistic combs in terms of their discriminability-distance.  In
the present paper we will restrict to finite-dimensional protocols,
with finite-number of rounds. The last assumption does not introduce
any practical limitation, since, in the real world one needs to put a
bound anyway to the lapse of time needed for the commitment. We will
anyway discuss also protocols with unbounded number of rounds in the
concluding section.
 
Before starting the main sections of the paper, we compare here the
present approach with that of the previous impossibility proof in Ref.
\cite{Kretschmann:2008p3022}. Ref.  \cite{Kretschmann:2008p3022}
treats the strategies as the preparation of a quantum register, and
classical and quantum communications are described in the Heisenberg
picture in the unified framework of C${}^*$-algebras. In the present
approach the C${}^*$-algebraic framework is avoided, by treating
classical histories as labels for sequences of quantum operations in
the Schr\"{o}dinger picture, and strategies are identified with
conditional quantum combs, which provide a direct mathematical
formulation. At this level, the differences are only in the
mathematical language, but two approaches are substantially
equivalent.  There are, however, conceptual differences, in which the
two approaches sensibly differ. The most relevant difference is the
notion of security, which in the present treatment is taken at the
strongest level, {\em i.e.} worst-case over all classical histories,
whereas in Ref.  \cite{Kretschmann:2008p3022} security was defined in
average. The present security notion is cryptographically the
strongest, corresponding to a priceless commitment bit. Another
important difference between the present approach and that of Ref.
\cite{Kretschmann:2008p3022} is a more general impossibility proof, in
which one can restrict the set of possible Bob's operations to a set closed under dilations.  In other words, we assume that Bob
is able to keep pure his quantum information and to perform arbitrary
quantum operations on his ancillae, whereas his operations on the
quantum systems exchanged during the commitment can be restricted by
arbitrary constraints. This makes the impossibility proof more
general, including {\em e.g.} the case of a Bob constrained by a
checking Alice.

The paper is organized as follows. In section \ref{s:proto} we review
the definition and the main features of a quantum protocol for bit
commitment, and define what a successful bit commitment protocol would
have to achieve.  In Section \ref{s:qcombs} we will briefly recall the
prerequisites about quantum combs, including the notion of quantum
tester, the discriminability-distance, the dilation of combs, and the
notion of conditional comb. The most important result of the section
is the dilation theorem for quantum combs, along with a continuity
theorem in terms of the discriminability-distance. In Section
\ref{s:mathform} we present the mathematical formulation of bit
commitment in terms of quantum combs, and state the impossibility
proof for protocols with bounded and unbounded number of rounds of
communication. The analysis will be based solely on the principles of
quantum mechanics, including classical physics, but not including
relativistic constraints, which are known to facilitate secure bit
commitment \cite{Ken99,Ken05}.  Section \ref{s:conc} concludes the
paper with some comments on the main results.

\section{What is a protocol}\label{s:proto}

A protocol regulates the exchange of messages between participants,
defining what are the honest strategies that they can adopt,
so that at every stage it is clear what type of message is expected
from the participants, although, of course, their content is not
fixed. The expected message types can be either classical or quantum
or a combination thereof. The number of classical states and the
dimension of the Hilbert spaces at a given step can depend on the
previously generated classical information.


\subsection{Phases of the Protocol}

\begin{figure}[h]
  \epsfig{file=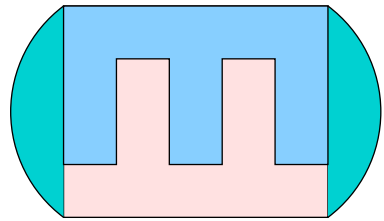,width=5cm}
  \caption{\label{f:third_party} The bit commitment protocol is
    two-party only, and trusted third parties are not allowed. Here in
    figure rounded portions represent examples of trusted third
    parties, {\em e.g.} the left one could be a trusted joint state,
    and the right one a trusted joint measurement. Another example of
    third party could be a third comb interlaced with Alice's and
    Bob's.}
\end{figure}

In any bit commitment protocol, we can distinguish two main phases.  The
first is the {\it commitment phase}, in which Alice and Bob exchange
classical and quantum messages in order to commit the bit.
Eventually, this phase can end either with a successful commitment, or
with an abort, in which the two parties irrevocably give up the purpose
of committing the bit (of course, in a well-designed protocol, if both parties are honest the
probability of abort should be vanishingly small). If no abort took
place, the bit value is considered to be committed to Bob but,
supposedly, concealed from him.  Since bit commitment is a two-party
protocol and trusted third parties are not allowed, the starting state
necessarily has to be originated by one of the two parties (see also
Fig.  \ref{f:third_party}).
Moreover, since we can always include in the protocol null steps (in which no
information, classical or quantum, is exchanged), without loss of
generality, we can restrict our attention to protocols that are
started by Bob.


The second phase is the {\it opening phase}.  In the case of abort
during the commitment, this is just a null step, whereas, in the case
of successful commitment, at the opening Alice will send to Bob some
classical or quantum information in order to to reveal the bit value.
Taking both Alice's message and his own (classical and quantum)
records, Bob will then perform a suitable {\it verification
  measurement}. His measurement will result in either a successful
readout of the committed bit, or in a failure, {\em e.g.} due to the
detection of an attempted cheat.  Again, in a well-designed protocol
the probability of failure should be vanishingly small.




\subsection{Conditions on Successful Protocols}

In the following we will denote by $a_0$ and $a_1$ two honest
strategies corresponding to the two bit values 0 and 1, respectively.
We call a protocol $\varepsilon$-{\it concealing} if, conditionally on
any history of classical communication, Bob cannot distinguish between
the strategies $a_0$ and $a_1$ (up to an error $\varepsilon$) before
Alice opens the commitment.  In general, of course, the probability of
a given history of classical communication depends on whether Alice
chooses $a_0$ or $a_1$. Since this dependence can be exploited by Bob to
infer the bit value, we must require that, no matter what strategy $b$
Bob uses, the conditional probability of $a_0$ given history $s$ never
differs from the probability of $a_1$ given history $s$ by more than
$\varepsilon$.  Note that this requirement must by satisfied even by
histories that end up in an abort, otherwise, by the sole fact that
the protocol aborted Bob could reliably infer the value of the bit.

We say that an Alice's strategy $a^\sharp$ is \emph{$\delta$-close} to
$a$ if, conditionally on any history of classical communication, Bob
cannot distinguish $a$ from $a^\sharp$ (up to an error $\delta$) at
any time, including the opening phase. Given two honest strategies $a_0$ and $a_1$, a
\emph{$\delta$-cheating} is a pair of strategies $a_0^\sharp$ and
$a_1^\sharp$, with the properties that \emph{i)} $a_i^\sharp$ is
$\delta$-close to $a_i$ for $i=0,1$ and \emph{ii)} Alice can turn
$a_0^\sharp$ into $a_1^\sharp$ with a local operation on her ancillae
after the end of the commitment phase.  In other words, the strategies
$a_0^\sharp$ and $a_1^\sharp$ are the same throughout the
commitment phase, and differ only by a local operation carried out
before the opening.    If no $\delta$-cheating strategy exists for
Alice, we call the protocol \emph{$\delta$-binding}.

\section{Prerequisites on quantum combs}\label{s:qcombs}

Here we briefly summarize the formalism of quantum combs and few
related results. In addition, this section contains the core result of this paper, namely the continuity theorem for the dilation of probabilistic quantum combs in terms of their discriminability-distance.  

\subsection{Choi-Jamio\l kowski operators  and link product}
A quantum operation (trace non-increasing CP-map) $\map C$ from states
on $\spc H_i$ to states on $\spc H_j$ is described by its Choi-Jamio\l kowski operator
\begin{equation}
  C = (\map C \otimes \map I_i)(|I_i\kk \bb I_i|) \in \Lin{ \spc H_j \otimes \spc H_i},
\end{equation}
where $\map I_i$ is the identity map on $\spc H_i$, and $|I_i\kk \in \spc H_i^{\otimes 2}$ is the
maximally entangled vector $|I_i\kk = \sum_n |n\>|n\>$, $\{|n\>\}$ orthonormal basis for $\spc H_i$. By the Choi's theorem, the map $\map C$ is
CP if and only if the Choi-Jamio\l kowski operator is positive (semidefinite).  In general, we will often exploit
the one-to-one correspondence between bipartite states in $|F\kk \in \spc H_j \otimes \spc H_i$ and
operators $F$ from $\spc H_i$ to $\spc H_j$ given by
\begin{equation}
|F\kk = (F \otimes I_i) |I_i\kk,
\end{equation}  
and the useful relation
\begin{equation}\label{mirrorket}
(F \otimes I_i) |I_i \kk = (I_j \otimes F^\tau) |I_j\kk,
\end{equation}
$F^\tau$ denoting the transposed of $F$ with respect to the orthonormal basis $\{|n\>\}$.  If $\map
C$ is a quantum operation from $\spc H_i$ to $\spc H_j$ and $\map D$ is a quantum operation from
$\spc H_j$ to $\spc H_k$, the Choi-Jamio\l kowski operator of the quantum operation $\map D\, \map C$, from $\spc
H_i$ to $\spc H_k$, resulting from the connection of $\map C$ and $\map D$ is given by the
\emph{link product} \cite{qca}
\begin{equation}\label{link} 
  D *
  C: = \Tr_{j} [(D \otimes I_i) (I_k \otimes C^{\tau_j}) ],
\end{equation} 
$\Tr_j$ and $\tau_j$ denoting partial trace and partial transpose on
$\spc H_j$, respectively.  A quantum operation $\map C$ is
trace-preserving ({\em i.e.} it is a channel) if and only if it satisfies
the normalization condition
\begin{equation} I_j * C \equiv \Tr_j[C] = I_i.
\end{equation}
Viewing quantum states as a special kind of channels (with
one-dimensional input space), Eq.  (\ref{link}) yields
\begin{equation}\label{normchannel}
\map C (\rho) = C * \rho = \Tr_i[ C (I_j \otimes
\rho^\tau )].
\end{equation}

\subsection{Quantum combs}

\begin{figure}[h]
\begin{center}
\setlength{\unitlength}{.45cm}
\begin{picture}(19,5.5)(0,0) 
  \put(0,2.75){\line(1,0){1}}
  \put(3,2.75){\line(1,0){.5}}
  \put(5.5,2.75){\line(1,0){.5}}
  \put(8,2.75){\line(1,0){.5}}
  \put(10,2.75){\line(1,0){.5}}
  \put(12.5,2.75){\line(1,0){.5}}
  \put(15,2.75){\line(1,0){.5}}
  \put(17.5,2.75){\line(1,0){1}}
  \put(1,0.5){\line(0,1){3}}
  \put(3,0.5){\line(0,1){3}}
  \put(1,0.5){\line(1,0){2}}
  \put(1,3.5){\line(1,0){2}}
  \put(6,0.5){\line(0,1){3}}
  \put(8,0.5){\line(0,1){3}}
  \put(6,0.5){\line(1,0){2}}
  \put(6,3.5){\line(1,0){2}}
  \put(15.5,0.5){\line(0,1){3}}
  \put(17.5,0.5){\line(0,1){3}}
  \put(15.5,0.5){\line(1,0){2}}
  \put(15.5,3.5){\line(1,0){2}}
  \put(10.5,.5){\line(0,1){3}}
  \put(12.5,.5){\line(0,1){3}}
  \put(10.5,.5){\line(1,0){2}}
  \put(10.5,3.5){\line(1,0){2}}
  \put(3,1.25){\line(1,0){3}}
  \put(8,1.25){\line(1,0){.5}}
  \put(10,1.25){\line(1,0){.5}}
  \put(12.5,1.25){\line(1,0){3}}

  \put(1.6,1.8){$\map C_0$}
  \put(6.7,1.8){$\map C_1$}
  \put(10.65,1.8){$\map C_{N-2}$}
  \put(15.65,1.8){$\map C_{N-1}$}
 
  \multiput(9,2.75)(0.15,0){4}
  {\line(1,0){0.05}}
  \multiput(9,1.25)(0.15,0){4}
  {\line(1,0){0.05}}
\end{picture}
\end{center}
\caption{\label{memch} $N$-comb: sequential network of $N$ quantum
  operations with memory. The network contains input and output
  systems (free wires in the diagram), as well as internal memories
  (wires connecting the boxes).}
\end{figure}
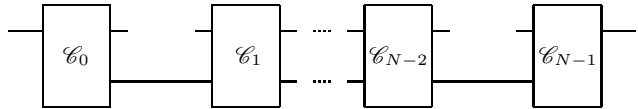

A quantum comb describes a sequential network of $N$ quantum
operations with memory $(\map C_k)_{k=0}^{N-1}$, with $N-1$ open slots in
which variable quantum operations can be inserted, as in Fig.
\ref{memch}.
The comb is in one-to-one correspondence with the the Choi-Jamio\l kowski operator
$R$ of the network, which can be computed as the link product of the
Choi-Jamio\l kowski operators $( C_k)_{k=0}^{N-1}$:
\begin{equation}
  R :=  C_{N-1} * \dots *  C_0. 
\end{equation}  Labelling the input (output)
spaces of $\map C_k$ as $\spc H_{2k}~ (\spc H_{2k+1})$, we have that
$R$ is a non-negative operator  on $\spc K:= {\bigotimes_{j=0}^{2N-1} \spc H_j}$.

For networks of channels the operator $R $ has to satisfy the
recursive normalization condition \cite{qca,watrous}
\begin{equation}\label{recnorm}
\Tr_{2k-1} [ R^{(k)}] = I_{2k-2} \otimes R^{(k-1)} \qquad k=1, \dots, N~
\end{equation} 
where $R^{(N)} := R$, $R^{(k)} \in \Lin{\bigotimes_{j=0}^{2k-1} \spc
  H_j} $, and $R^{(0)} =1$.  Moreover, one has the
characterization \cite{qca,comblong}
\begin{Theo}\label{prorea}
  Any positive operator $R$ satisfying Eq. (\ref{recnorm}) is the Choi-Jamio\l kowski
  operator of a sequential network of $N$ channels. Any positive operator $R'$
  such that $R' \le R$ is the Choi-Jamio\l kowski operator of a sequential network of $N$
  quantum operations.
\end{Theo} 
We call a quantum comb $R$ satisfying Eq.  (\ref{recnorm})
\emph{deterministic}, and a comb $R' \le R$ \emph{probabilistic}.

\subsection{Dilation of quantum combs}

By Stinespring-Kraus-Ozawa theorem\cite{Stinespring,Kraus,Ozawa}, any quantum operation $\map C$
from states on $\spc H_i$ to $\spc H_j$ can be dilated to an isometric map followed by a
post-selection on an ancilla
\begin{equation}
\begin{split}
  \map C (\rho) &= \Tr_A [(I \otimes P_A) V \rho V^\dag ]\\
  & = \Tr_A [K \rho K^\dag]  \qquad K = (I
  \otimes P_A) V,
\end{split}
\end{equation}  
with $V$ isometry from $\spc H_i$ to $\spc H_j\otimes \spc H_A$, and $P_A$ orthogonal projector on a subspace of the ancilla space $\spc H_A$.

We refer to the single-Kraus map
\begin{equation}\label{single-kraus}
\widetilde{\map C} (\rho):= K \rho K^\dag
\end{equation}
as to a \emph{dilation} of the quantum operation $\map C$.  In terms of Choi-Jamio\l kowski
operators, one has
\begin{equation}
C = \Tr_A [\widetilde C]\in\Lin{\spc H_j\otimes \spc H_i},  
\end{equation}
where 
$\widetilde C = |K\kk \bb K |$ is the Choi-Jamio\l kowski operator of the dilation. A
(minimal) dilation of the quantum operation $\map{C}$ has ancilla space $\spc H_A\simeq\Supp (C)
\subseteq \spc H_j \otimes \spc H_i := \spc H_{ij}$, and Choi-Jamio\l kowski operator
\begin{equation}
  \widetilde C  = |C^{\frac 1 2} \kk \bb C^{\frac 1 2}|  \in \Lin{\spc H_{ij} \otimes \spc H_A}.
\end{equation}
In particular, when the quantum operation is a quantum channel also its dilation is a
channel---$\widetilde{\map C} (\rho) =V \rho V^\dag$, $V$ isometry---with the Choi-Jamio\l kowski
operator satisfying the normalization condition $\Tr_{A,j} [\widetilde C] = I_i$.

Since a quantum comb $R \in \Lin {\spc K}$  with $\spc K = \bigotimes_{j =
  0}^{2N-1} \spc H_j$ represents a sequential network of quantum
operations, one can always obtain a dilation of the comb by
dilating each quantum operation in the network. A useful dilation
of $R$ is given by
\begin{equation}
  \widetilde R = |R^{ \frac 1 2 } \kk \bb R^{\frac 1 2} | \in \Lin{\spc K \otimes \spc H_A},
\end{equation} 
where $\spc H_A\simeq\Supp (R) \subseteq \spc K$.
The dilation $\widetilde R$ has the following interpretation:
$\widetilde R$ is a quantum comb acting on the Hilbert spaces $\left(\widetilde {\spc
H}_j\right)_{j=0}^{2N-1}$, where $\widetilde{\spc H}_{2N-1} := \spc
H_{2N-1} \otimes \spc H_A$, and $\widetilde {\spc H}_k = \spc H_k$ for
$k< 2N-1$. Therefore, by Theorem \ref{prorea} it represents a sequence of $N$ quantum operations
with memory $(\widetilde{\map C}_k)_{k=0}^{N-1}$. Tracing out the ancilla space $\spc H_A$ in the output
$\widetilde{\spc H}_{2N-1} = \spc H_{2N-1} \otimes \spc H_A$ of the
last quantum operation $\widetilde{\map C}_{N-1}$, one then obtains back the
original network
\begin{equation}
R = \Tr_A \left[\widetilde R \right].
\end{equation}
Note that only the ancilla space $\spc H_A$ in the output of the last
quantum operation appears in the dilation $\widetilde R$.

For quantum states it is known that the purification is unique up to
partial isometries on the ancilla spaces. For quantum combs one has
the straightforward extension:
\begin{Prop}
  Let $R \in \Lin{\spc K}$, $\spc K= \bigotimes_{j=0}^{2N-1} \spc H_j$ be a
  quantum comb. Let $\widetilde R \in \Lin{\spc K \otimes \spc H_A}$
  and $\widetilde R' \in \Lin{\spc K\otimes \spc H_{A'}}$ be two
  dilations of $R$, i.e. $\widetilde R$ and $\widetilde R' $ are both
  non-negative rank-one operators such that
\begin{equation}
  \Tr_{A} \left[ \widetilde R \right] = \Tr_{A'} \left[ \widetilde R'\right].
\end{equation}  
Then there exists a partial isometry $W $ from $\spc H_A$ to $\spc H_{A'}$ such that
\begin{equation}
\begin{split}
&  \widetilde R' = (I \otimes W) \widetilde R (I \otimes W^\dag),\\  
&\widetilde R  = (I \otimes W^\dag) \widetilde R' (I \otimes W),
\end{split}
\end{equation}
$I$ denoting the identity on $\spc K$.
\end{Prop}
For the application to bit commitment it is crucial to note that all dilations of a comb can be
obtained by just applying a partial isometry $W$ on the \emph{last output system}. An obvious
consequence of the above fact is:
\begin{Cor}
Let  $R \in \Lin{\spc K}, \spc K= \bigotimes_{j=0}^{2N-1} \spc H_j$ be a quantum comb.   If $\widetilde R \in \Lin{\spc K \otimes \spc H_A}$ and $\widetilde R'\in \Lin{\spc K \otimes \spc H_{A'}}$ are two dilations of $R$, then there exist
  two quantum channels $\map E$ from states on $\spc H_A$ to states on $\spc H_{A'}$ and $\map F$
  from states on $\spc H_{A'}$ to states on $\spc H_A$ such that
\begin{equation}
\begin{split}
  & \widetilde R' = \left(\map I \otimes \map E\right)
  (\widetilde R) = E * \widetilde R\\& \widetilde R = \left(\map I \otimes
    \map F\right) (\widetilde R') = F * \widetilde R',
\end{split}
\end{equation}
$\map I$ denoting the identity map on $\spc K$, and $E$ and $F$ being the Choi-Jamio\l kowski operators of the channels $\map E$
and $\map F$, respectively.
\end{Cor}
This means that one can switch from one dilation to another just by
performing some physical transformation on the ancilla in output  of the last quantum operation of the comb.  As we will see in the
following, in a bit commitment protocol this implies that Alice can
delay her choice of the bit to the last moment before the opening.

\subsection{Quantum testers}

\begin{figure}[h]
\setlength{\unitlength}{.5cm}
\begin{center}
\begin{picture}(19.5,5.5)(.7,0) 
 \put(3,2.75){\line(1,0){.5}}
  \put(5.5,2.75){\line(1,0){.5}}
  \put(8,2.75){\line(1,0){.5}}
  \put(10,2.75){\line(1,0){.5}}
  \put(12.5,2.75){\line(1,0){.5}}
  \put(15,2.75){\line(1,0){.5}}
  \put(3,2){\oval(4,3)[l]}
  \put(3,0.5){\line(0,1){3}}

  \put(6,0.5){\line(0,1){3}}
  \put(8,0.5){\line(0,1){3}}
  \put(6,0.5){\line(1,0){2}}
  \put(6,3.5){\line(1,0){2}}
  \put(3.5,2){\line(0,1){3}}
  \put(5.5,2){\line(0,1){3}}
  \put(3.5,2){\line(1,0){2}}
  \put(3.5,5){\line(1,0){2}}
  \put(13,2){\line(0,1){3}}
   \put(15,2){\line(0,1){3}}
  \put(13,2){\line(1,0){2}}
  \put(13,5){\line(1,0){2}}
  \put(10.5,.5){\line(0,1){3}}
  \put(12.5,.5){\line(0,1){3}}
  \put(10.5,.5){\line(1,0){2}}
  \put(10.5,3.5){\line(1,0){2}}
  \put(3,1.25){\line(1,0){3}}
  \put(8,1.25){\line(1,0){.5}}
  \put(10,1.25){\line(1,0){.5}}
  \put(12.5,1.25){\line(1,0){3}}
  \put(5.5,4.25){\line(1,0){2.5}}
   \put(10.5,4.25){\line(1,0){2.5}}
  \put(15.5,2){\oval(4.3,3)[r]}
   \put(15.5,0.5){\line(0,1){3}} 

  \put(1.6,1.8){$\rho_0$}
  \put(6.5,1.8){$\map D_1$}
  \put(10.8,1.8){$\map D_{N-1}$}
  \put(4.3,3.3){$\map C_0$}
  \put(13.4,3.3){$\map C_{N-1}$}
 \put(16.2,1.8){$P_i$}
  \multiput(9,2.75)(0.15,0){4}
  {\line(1,0){0.05}}
  \multiput(9,1.25)(0.15,0){4}
  {\line(1,0){0.05}}
  \multiput(9,4.25)(0.15,0){4}
  {\line(1,0){0.05}}
\end{picture}
\caption{\label{tester} Testing a network of $N$ quantum operations
  $(\map C_k)_{k=0}^{N-1}$. The tester consists in the preparation of
  an input state $\rho_0$, followed by quantum operations $\{\map D_1,
  \dots, \map D_{N-1}\}$, and a final measurement $\{P_i\}$.}
\end{center}
\end{figure}
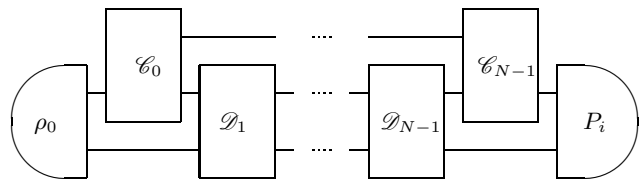

A tester represents a quantum network starting with a state
preparation and finishing with a measurement.  When such a network is
connected to a network of $N$ quantum operations as in Fig.
\ref{tester}, the output is a measurement outcome $i$ with probability
$p_i$. In a bit commitment protocol, a dishonest Bob will perform a
tester to distinguish Alice's strategies before the opening.

Mathematically, the tester is the collection of Choi-Jamio\l kowski operators
$\{T_i\}$ given by
\begin{equation}
  T_i := P_i * D_{N-1} * \dots * D_1 * \rho_0,
\end{equation} 
where $(D_k)_{k=1}^{N-1}$ are the Choi-Jamio\l kowski operators of the quantum operations $(\map
D_k)_{k=1}^{N-1}$ in Fig. \ref{tester}. 
If the sum over all outcomes $T= \sum_i T_i$ is a deterministic comb, we
call the tester \emph{normalized}.

When the tester is connected to a quantum network $R$, the probability of the outcome $i$ is
\begin{equation}
  p_i = T_i * R = \Tr \left[  T_i^{ \tau}  R \right],
\end{equation}
which is nothing but the Born rule, for quantum networks rather than
states. Notice that one can include the transpose in the definition
of the tester, thus getting the familiar form of the Born rule $p_i =
\Tr[ T_i R ]$.  However, here we preferred to write probabilities in
terms of the combs $R$ and $T_i$ of the measured and
measuring networks, respectively, thus making explicit that the Born
rule is nothing but a particular case of link product, the transpose
appearing as the signature of the linking of two networks.

For a deterministic comb $R$ and a normalized tester $\{T_i\}$ one has the normalization of the total
probability:
\begin{equation}\label{normprob}
  \sum_i p_i  = \sum_i  \Tr[T^\tau_i  R] =1.
\end{equation}
In general, if one considers sub-normalized testers, one has
\begin{equation}
  \sum_i p_i  = \sum_i  \Tr[T^\tau_i  R] =p\le1.
\end{equation}

In the following we will call $T=\sum_i T_i$ {\em tester operator}.

\begin{Prop}[Decomposition of testers \cite{memorydisc}]\label{retest}
  Let $ T \in \Lin{\spc K}, \spc K=\bigotimes_{j=0}^{2N-1} \spc H_j$ be the tester operator of the quantum tester $\{T_i \}$.  Let $\spc H_B$ be
  the ancilla space $\spc H_B \simeq \Supp \left( T \right)\subseteq \spc K$, and $\widetilde T$ be the dilation
  given by
\begin{equation}
  \widetilde T  = | T^\frac 1 2 \kk \bb T^{\frac 1 2} | \in \Lin{\spc H \otimes \spc H_B}.
\label{purtest}
\end{equation} 
Then, one has the identity
\begin{equation}\label{sqrtt}
  \widetilde T * R = \left[ T^{\tau} \right]^{\frac 1 2} R \left[ T^{\tau} \right]^{\frac 1 2}.
\end{equation}
Moreover, the probabilities of outcomes $p_i = T_i * R$ are given by
\begin{equation}\label{testerdeco}
  p_i = P_i * \widetilde T * R, 
\end{equation}  
where $\{P_i\}$ is the POVM on $\spc H_B$ defined by
\begin{equation}
  P_i = T^{-\frac 1 2} ~ T_i ~ T^{-\frac 1 2},
\end{equation}
$T^{-1/2}$ being the inverse of $T^{1/2}$ on its support.
\end{Prop}
\Proof  Checking Eq. (\ref{sqrtt}) is immediate using Eq. (\ref{mirrorket})
\begin{equation}
\begin{split}
\widetilde T * R &= \Tr_{\spc H} [(R^{\tau} \otimes I_B) |T^{\frac 1 2} \kk \bb T^{\frac 1 2}|]\\
&= \left[ T^{\tau} \right]^{\frac 1 2} R \left[ T^{\tau}
\right]^{\frac 1 2}.
\end{split}
\end{equation}  
Regarding Eq. (\ref{testerdeco}), one has $p_i = T_i * R = \Tr
[T_i^\tau R] = \Tr[ \left[ T^{\tau} \right]^{\frac 1 2} P_i^\tau
\left[ T^{\tau} \right]^{\frac 1 2} R] = \Tr[P_i^\tau (\widetilde T *
R)] = P_i * \widetilde T * R $.  \qed

The interpretation of the above result is the following realization scheme for the tester $\{T_i\}$: 
\begin{itemize}
\item realize the quantum network $\widetilde T$ and connect it with the measured network $R$ 
\item conditionally on the given history of classical information corresponding to $\widetilde T$,
  perform the POVM $\{P_i\}$ on the ancilla state $\rho = \widetilde T * R$.
\end{itemize}
\subsection{Discriminability of combs}\label{subsect:discriminability}

Proposition \ref{retest} reduces any measurement on quantum network
$R$ to a measurement on a suitable (sub-normalized) state $\rho =
\widetilde T * R$, which is obtained by connecting the input comb $R$
with a suitable comb $\widetilde T$ corresponding to the dilation
of Eq.~\eqref{purtest}. In particular, it reduces the discrimination
of two networks $R_0$ and $R_1$ to the discrimination of two output
states 
\begin{equation}
\rho^{(i)}_{T} = \widetilde T * R_i  = \left[T^{\tau} \right]^{\frac 1 2} R_i \left[ T^\tau \right]^{\frac 1 2} \qquad i =0,1.
\end{equation}  This allows for the definition of an operational distance between networks
\cite{memorydisc}, whose meaning is directly related to statistical
discriminability
\begin{equation}\label{norm}
\begin{split}
  \left|\! \left | R_1 - R_0 \right| \!\right|_\mathrm{op} & := \sup_T  |\!|\rho^{(1)}_{T} - \rho^{(0)}_{T}  |\!|_1 \\ 
&=\sup_T |\!|
  \widetilde T * (R_1 -R_0) |\!|_1\\& = \sup_{T} \left| \! \left |
      \left[T^\tau\right]^{\frac 1 2} (R_1 -R_0) \left[T^\tau\right]^{\frac 1 2} \right|\!\right|_1,
\end{split}
\end{equation}  
where twe supremum is taken over the set of all tester operators $T =
\sum_i T_i$, and $|\!| A |\!|_1 = \Tr |A|$.  Remarkably, the above
norm can be strictly greater than the cb-norm of the difference $\map
R_1 - \map R_0$ of the two multipartite channels \cite{memorydisc}, since a sequential scheme such as that in Fig. \ref{tester} can
achieve a strictly better discrimination than a parallel scheme where
a multipartite entangled state is fed in the unknown channel.

When the tester $T$ and the combs $R_i$ are
probabilistic (namely correspond to networks of quantum operations)
the states $\rho^{(i)}_{T} =\widetilde T * R_i$ are generally
sub-normalized, {\em i.e.} $\Tr[\rho^{(i)}_{T}] \le 1$.  In this case,
the sole fact that the sequences of quantum operations represented by
$\widetilde T$ and $R_i$ took place helps in discriminating between $R_0$ and
$R_1$.  To be concrete, consider the scenario in which $R_0$ and $R_1$ have flat prior
probabilities $\pi_0 = \pi_1 = 1/2$.  The probability that the sequence of operations represented by
$\widetilde T$ and $R_i$ takes place is then given by $p(\widetilde T, R_i) = \Tr[\rho^{(i)}_T]/2$.  Since this probability
depends on $i$, upon knowing that the sequence of quantum operations $\widetilde T$ took place the initial flat
prior must be updated to 
\begin{equation}
\pi_i' = p(R_i|\widetilde T) = \frac{p(\widetilde T,R_i)}{p(\widetilde T)} =\frac{\Tr[\rho^{(i)}_T]}{\Tr[\rho_T^{(0)}+\rho_T^{(1)}]}. 
\end{equation}   
The discrimination is now between the two conditional states $\bar
\rho^{(i)}_{T}:=\frac{\rho^{(i)}_{T}}{\Tr[\rho^{(i)}_{T}]}$ with prior probability $\pi'_i$, 
$i=0,1$. Therefore, the maximum success probability is given by
\begin{equation}
\begin{split}
p_{succ}=&\frac{1}{2}\left(1+\left\|\pi'_0\bar\rho_T^{(0)}-\pi'_1\bar\rho_T^{(1)}\right\|_1\right)\\=&
\frac{1}{2}\left(1+\frac{\left\|\rho_T^{(0)}-\rho_T^{(1)}\right\|_1}{\Tr[\rho_T^{(0)}+\rho_T^{(1)}]}\right).
\end{split}
\end{equation}
Accordingly, we introduce the comb discriminability ``distance''
\begin{equation}
\begin{split}
  d(R_1,R_0)&:= {\sup_T}' \frac{\left\|\rho^{(1)}_T  -\rho^{(0)}_T \right\|_1}{\Tr\left[\rho^{(1)}_T  +\rho^{(0)}_T\right]}
\\
  & ={\sup_T}'\frac{\left\|\widetilde T*(R_1-R_0)\right\|_1}{\Tr[\widetilde
      T*(R_1+R_0)]}\\ & = {\sup_T}' \frac{ \left\|\left[T^\tau\right]^{\frac 1 2}(R_1
        -R_0)\left[T^\tau\right]^{\frac 1 2}\right\|_1}{\Tr[T^\tau
          (R_1+R_0)]},
\end{split}
\end{equation}
where $\sup'$ (and consistently $\inf'$) denotes the supremum
(infimum) restricted to the tester operators $T$ such that $\Tr[T^\tau
(R_0+R_1)]>0$.  Here and in the following, we use the word "distance"
informally, although for probabilistic combs the function $d$ is just
a semi-metric, namely the triangular inequality does not hold ({\em
  e.g.} consider the states $\rho=1/2(|0\>\<0|)$,
$\sigma=1/2(|1\>\<1|)$, and $\tau=I/2$, for which
$d(\rho,\sigma)>d(\rho,\tau)+d(\tau,\sigma)$).
 
\subsubsection{Discriminability with a restricted set of testers}
The comb distance quantifies the performances of the best scheme among
all possible sequential schemes one can use to discriminate between
two quantum networks. However, in a bit commitment protocol the set of
schemes that Bob can actually use for discrimination may be limited by
several factors.  For example, Alice could perform random checks
during the commitment phase in order to force Bob to use a quantum
network that is close to the one prescribed by the honest strategy.
We will therefore define optimal discrimination between $R_0$ and
$R_1$ relatively to a restricted set $\mathsf T$ of tester
operators that can actually occur in the protocol, thus introducing
the discriminability ``distance''
\begin{equation}\label{e:condis}
\begin{split}
  \left. d (R_1,R_0)\right|_{\mathsf T} &:=
{\sup_{T\in\mathsf T}}' \frac{\left\|\rho^{(1)}_T  -\rho^{(0)}_T \right\|_1}{\Tr\left[\rho^{(1)}_T
      +\rho^{(0)}_T\right]}\\&= 
{\sup_{T\in\mathsf T}}' \frac{ \left\|\left[T^\tau\right]^{\frac 1 2}(R_1
        -R_0)\left[T^\tau\right]^{\frac 1 2}\right\|_1}{\Tr[T^\tau
          (R_1+R_0)]}.
\end{split}
\end{equation}
The restriction of possible testers to a general set implies
that the distance defined in Eq.~\eqref{e:condis} does not satisfy the
property $d(x,y)=0\Rightarrow x=y$.
\begin{Lem}\label{l:trc-dist} The discriminability distance in Eq. (\ref{e:condis}) is monotone under
  the application of a channel on the ouput spaces, namely
\begin{equation}
  \left. d ((\map{C}\otimes\map{I}_{in})R_1,(\map{C}\otimes\map{I}_{in})R_0)\right|_{\mathsf T} \leq  \left. d (R_1,R_0)\right|_{\mathsf T}.
\end{equation}
\end{Lem}
\Proof Use monotonicity of trace-distance and the fact that the map $\map{C}$ is trace-preserving.\qed

\subsection{Continuity of dilation} 

We now prove that if two quantum combs $R_0$ and $R_1$ are close to
each other then there exist two dilations $\widetilde R_0$ and
$\widetilde R_1$ that are close with respect to the discriminability
distance.  Such continuity theorem replaces the Stinespring's
continuity theorem \cite{Kretschmann:2008p3022} used in the previous
(C${}^*$-algebraic) impossibility proof of Ref. \cite{werqbc}.


\begin{Lem}[Continuity of dilation]\label{conticondcomb}
  Let $R_0, R_1 \in \Lin {\spc K}$ be two quantum combs, $\widetilde
  R_i = |R_i^{\frac 1 2} \kk \bb R_i^{\frac 1 2}| \in \Lin{\spc K
    \otimes \spc H_A } , \spc H_A \simeq \spc K$ be two dilations,
  and $\mathsf T \subseteq \Lin {\spc K}$ be an arbitrary set of
  tester operators $T$. The following bound holds
\begin{align}
  \inf_{\mathcal P} \left. d\left(\widetilde R_1,(\mathcal I\otimes\mathcal P)(\widetilde
      R_0)\right)\right|_{{\mathsf T}\otimes I}
  \le \sqrt{2\left.d (R_0, R_1)\right|_{\mathsf T}}
\end{align}
where ${\mathsf T}\otimes I\:=\{T\otimes I|T\in\mathsf T\}$ and the
infimum is taken over the set of random unitary channels $\mathcal P
(\rho)=\sum_k p_k ~ U_k \rho U_k^\dag$ acting on the ancilla $\spc H_A
$.
\end{Lem}
\Proof If we define
\begin{equation} 
  \widetilde{\Delta}_{U_k}:= \frac{\widetilde R_{1}-(I \otimes U_k)\widetilde
    R_{0} (I \otimes U_k^\dag)}{\Tr[(\widetilde R_0+\widetilde R_1)*T]}
  ,
\end{equation}
we have
\begin{equation}
  \inf_\mathcal P \left. d \left(\widetilde R_1,(\mathcal I\otimes\mathcal P)(\widetilde
      R_0)\right)\right|_{{\mathsf T}\otimes I}=\inf_\mathcal P{\sup_{T\in\mathsf T}}'\left\|\sum_k  p_k\widetilde T*\widetilde\Delta_{U_k}\right\|_1.
\end{equation}
The triangle inequality for the trace-norm yields
\begin{align}
  &\left|\!\left|\sum_k p_k \widetilde T*\widetilde\Delta_{U_k}\right|\!\right|_1 \leq\sum_k p_k\left|\!\left|\widetilde T*\widetilde\Delta_{U_k}\right|\!\right|_1.
\label{fibound}
\end{align}
Moreover, exploiting Eq. (\ref{sqrtt}) we can write
\begin{equation}
  \left\| \widetilde T*\widetilde{ \Delta}_{U_k} \right\|_1= \left\|
  \Psi_{T,I}^{(1)}- \Psi^{(0)}_{T,U_k} \right\|_1,
\label{tracedist}
\end{equation} 
where $\Psi^{(0)}_{T, I}$ and $\Psi^{(1)}_{T, U_k}$ are defined by  
\begin{equation}
\begin{split}\label{purvec}
\Psi_{T,C}^{(i)} &:= |\Psi_{T,C}^{(i)} \kk \bb \Psi^{(i)}_{T,C}|\\
|\Psi_{T,C}^{(i)}\kk &:= \frac{( [T^\tau]^{\frac 1 2} \otimes C ) |R^{\frac 1 2}_{i} \kk}{\sqrt{\Tr[(R_0+R_1)T^\tau]}},
\end{split}
\end{equation}
for  $C\in\Lin{\spc H_A}$ any contraction. Using the bound
\begin{equation}\label{elementary}
\begin{split}
  |\!|  |\psi\>\<\psi| &- |\varphi\>\<\varphi|  |\!|_1^2  =  (|\!| \psi |\!|^2 + |\!| \varphi |\!|^2)^2   -4  | \<\psi | \varphi \> |^2  \\
 \le & (|\!| \psi |\!| + |\!| \varphi |\!|)^2 (|\!| \psi |\!|^2 + |\!|
 \varphi |\!|^2 - 2 | \< \psi | \varphi \> | ) ,
\end{split}
\end{equation}
which for $\| \psi \|^2+\| \varphi \|^2=1$ becomes
\begin{equation}\label{element1}
  |\!|  |\psi\>\<\psi| - |\varphi\>\<\varphi|  |\!|_1^2 \leq 2(1-2| \< \psi | \varphi \> | ) ,
\end{equation}
we obtain
\begin{equation}
  |\!| \widetilde T*\widetilde \Delta_{U_k} |\!|_1\leq   \sqrt 2 \left (1-2|\bb \Psi^{(1)}_{T,I}|\Psi^{(0)}_{T,U_k}\kk| \right)^\frac12.
\end{equation}
Then, by Jensen's inequality we have the following bound
\begin{equation}
\begin{split}
  & \sqrt 2\inf_\mathcal P {\sup_{T\in\mathsf T}}' \sum_k p_k\left (1-2|\bb \Psi^{(1)}_{T,I}|\Psi^{(0)}_{T,U_k}\kk| \right)^\frac12\le\\
 & \sqrt 2\inf_{\mathcal P}{\sup_{T\in\mathsf T}}' \left (1-2\sum_k p_k \left|\bb \Psi^{(1)}_{T,I} |\Psi^{(0)}_{T,U_k}\kk\right|  \right)^\frac12\le\\
  & \sqrt 2\inf_{\mathcal P}{\sup_{T\in\mathsf T}}' \left (1-2\left|\sum_k p_k\bb \Psi^{(1)}_{T,I} |\Psi^{(0)}_{T,U_k}\kk\right| \right)^\frac12=\\
 & \sqrt 2 \inf_{\mathcal P} {\sup_{T\in\mathsf T}}' \left(1- 2\left|\bb
    \Psi_{T,I}^{(1)} | \Psi^{(0)}_{T, C} \kk\right|
  \right)^{\frac 1 2}\leq\\
  & \sqrt 2 \inf_{\mathcal P} {\sup_{T\in\mathsf T}}' \left(1- 2{\rm Re}\bb\Psi_{T,I}^{(1)} | \Psi^{(0)}_{T, C} \kk\right)^{\frac 1 2} ,
\end{split}
\end{equation}
where $C$ is the contraction $C =\sum_k p_k U_k$. Let us define by
$\mathsf C$ the compact convex set of all contractions $C = \sum_k p_k U_k$, and
define the following function on $\mathsf C\times\mathsf T$
\begin{equation}
  f(C,T):= \mathrm{Re}  \bb \Psi^{(1)}_{T,I}  |\Psi^{(0)}_{T,C} \kk   ,
\label{functs}
\end{equation}
In Appendix \ref{app} we use Sion's minimax theorem of Ref. \cite{fan} to prove the identity
\begin{equation}\label{minimaxbound}
  {\inf_{T\in \mathsf T}}'\sup_{C\in\mathsf C} f(C,T)=\sup_{C\in\mathsf C} {\inf_{T\in \mathsf
      T}}' f(C,T). 
\end{equation}
The chain of inequalities proved until now gives
\begin{align}
  \inf_\mathcal P &\left.d \left(\widetilde R_1,(\mathcal I\otimes\mathcal P)(\widetilde R_0)\right) \right|_{\mathsf T }\le \sqrt 2 (1-2\sup_C{\inf_{T\in\mathsf T}}'f(C,T))^\frac12\nonumber\\
  &= \sqrt 2\left (1-2{\inf_{T\in\mathsf T}}' \sup_C f(C,T) \right)^\frac12=\\
  &\le \sqrt 2\left (1-2{\inf_{T\in\mathsf T}}' \sup_U f(U,T) \right)^\frac12,
\end{align}
where we substituted the supremum over contractions $C= \sum_k p_k
U_k$ with the supremum over unitaries $U$, since the function $f(T,C)$
is linear in $C$.  Moreover, we have
\begin{equation}
\begin{split} 
&\sup_U  f (T, U)  = \sup_U\mathrm{Re}\bb \Psi_{T,I}^{(0)} |I\otimes U |\Psi^{(1)}_{T,I} \kk \\
=&\sup_U  \bb \Psi_{T,I}^{(0)} |I\otimes U |\Psi^{(1)}_{T,I} \kk
=\frac{F\left(\rho^{(1)}_T,\rho^{(0)}_T \right)}{\Tr[\rho^{(1)}_T+\rho^{(0)}_T]}, 
\end{split}
\end{equation}
where $\rho^{(i)}_T$ $i=0,1$ denote the unnormalized states $\rho^{(i)}_T:=\left[T^{\tau} \right]^{\frac 1
  2} R_i \left[ T^\tau \right]^{\frac 1 2}$ and $F (\rho, \sigma) = \sup_U \Tr[\rho^{\frac 1 2} U \sigma^{\frac 
  1 2} ]$ is the Uhlmann fidelity. Finally, we can use the Bures-Alberti-Uhlmann bound
\begin{equation}
\Tr[\rho+\sigma]-2F(\rho,\sigma)\leq \|\rho-\sigma\|_1
\end{equation}
to obtain
\begin{align}
  \inf_\mathcal P& \left. d\left(\widetilde R_1,(\mathcal I\otimes\mathcal P)(\widetilde R_0)\right)
  \right|_{\mathsf T} \le\\ &\sqrt 2 {\sup_{T\in\mathsf T}}' \left(1-2\frac{F\left(\rho^{(1)}_T,\rho^{(0)}_T \right)}{\Tr\left[\rho_T^{(0)} +\rho_T^{(1)}\right]}\right)^{\frac 1 2}\nonumber\\
  & \le   {\sup_{T\in\mathsf T}}' \sqrt { \frac{2|\!|\rho^{(1)}_T -\rho^{(0)}_T |\!|_1}{
\Tr\left[\rho_T^{(0)} +\rho_T^{(1)}\right]}}=  ~\sqrt{2\left. d (R_1,R_0) \right|_{\mathsf T}}.
\end{align}
\qed
\subsection{Conditional quantum combs\label{s:ccombs}}
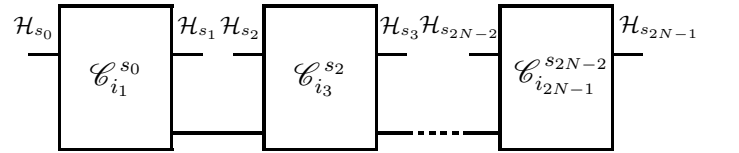
\begin{figure}[h]
\begin{center}
\setlength{\unitlength}{1300sp}
\begin{picture}(11898,2766)(-164,-2794)
\thicklines
{\put(4501,-2761){\framebox(2100,2700){}}}
{\put(9001,-2761){\framebox(2100,2700){}}}
{\put(  1,-961){\line( 1, 0){600}}}
{\put(3901,-961){\line( 1, 0){600}}}
{\put(2701,-961){\line( 1, 0){600}}}
{\put(8401,-961){\line( 1, 0){600}}}
{\put(11101,-961){\line( 1, 0){600}}}
{\put(2701,-2461){\line( 1, 0){1800}}}
{\put(6601,-2461){\line( 1, 0){600}}}
{\put(6601,-961){\line( 1, 0){600}}}
{\put(8401,-2461){\line( 1, 0){600}}}
{\multiput(7201,-2461)(218.18182,0.00000){6}{\line( 1, 0){109.091}}}
\put(1151,-1761){\makebox(0,0)[lb]{\large$\map{C}^{\,s_0}_{i_1}$}}
\put(5051,-1761){\makebox(0,0)[lb]{\large$\map{C}^{\,s_2}_{i_3}$}}
\put(9251,-1761){\makebox(0,0)[lb]{\large$\map{C}^{\,s_{2N-2}}_{i_{2N-1}}$}}
\put(-300,-661){\makebox(0,0)[lb]{$\spc H_{s_0}$}}
\put(2831,-661){\makebox(0,0)[lb]{$\spc H_{s_1}$}}
\put(3641,-661){\makebox(0,0)[lb]{$\spc H_{s_2}$}}
\put(6691,-661){\makebox(0,0)[lb]{$\spc H_{s_3}$}}
\put(7441,-691){\makebox(0,0)[lb]{$\spc H_{s_{2N-2}}$}}
\put(11251,-691){\makebox(0,0)[lb]{$\spc H_{s_{2N-1}}$}}
{\put(601,-2761){\framebox(2100,2700){}}}
\end{picture}
\caption{Sequence of quantum operations depending on previously
  exchanged classical information. Here $i_{2k-1}$ is the outcome of
  the $k$-th quantum operation, and $s_{l} = i_0 i_1 \dots i_{l}$ is
  the history of classical information available at step $l$.  The
  collection of all sequences corresponding to all possible classical
  histories is the conditional comb.  \label{f:ccomb}}
\end{center}
\end{figure}

A general two party protocol entails the exchange of both quantum systems
and of classical information, which is in principle openly known.
Therefore, the strategy of a party will result in a sequence of
quantum operations  $\map{C}^{\,s_{2k-2}}_{i_{2k-1}}$,  $k=1,2,\ldots,N$, 
as in Fig. \ref{f:ccomb}. Here the index $i_{2k-1}$ denotes the
outcome of the quantum operation, and the string $s_l$ represents the
full history of classical information exchanged before the occurrence of the
operation, namely $s_l=i_0i_1\ldots i_l$, with $i_{2k-2}$ representing
the input classical information at step $k$.
For example, if the comb in Fig. \ref{f:ccomb} represents Alice's
strategy in a two-party protocol with Alice's and Bob's combs
interlaced as in Fig. \ref{f:interlacedqcombs}, it describes the
following situation: Alice receives from Bob the classical information
$i_0\equiv s_0$ along with a quantum system with Hilbert space $\spc
H_{s_0}$. Then she performs the instrument
$\map{I}^{s_0}=\{\map{C}^{s_0}_j\}$ obtaining the outcome $j=i_1$.
After that she sends to Bob the outcome along with a quantum system with
Hilbert space $\spc H_{s_1}$ with $s_1=i_0i_1$. The normalization of
the instrument is
\begin{equation}
\sum_{i_1}\Tr_{\spc H_{s_1}}[\map{C}^{s_0}_{i_1}(\rho)]=\Tr[\rho],\quad\forall s_0,
\forall\rho\in\Lin{\spc H_{s_0}},
\end{equation}
which, in terms of Choi-Jamio\l kowski operators reads
\begin{equation}
\sum_{i_1}  \Tr_{\spc H_{s_1}}  [ C^{s_0}_{i_1}] = I_{s_0} \qquad \forall s_0.
\end{equation}
At the next step Alice receives from Bob the classical
information $i_2$ along with a quantum system with Hilbert space $\spc
H_{s_2}$, which depends on $s_2=i_0i_1i_2$. Then she performs the
instrument $\map{I}^{s_2}=\{\map{C}^{s_2}_j\}$ obtaining the outcome
$j=i_3$, and so on.  By linking the Choi-Jamio\l kowski operators of
all quantum operations, one obtains a family of probabilistic combs
$\{R_{s_{2N-1}}\}$ satisfying the normalization conditions
\begin{equation}\label{e:truegame}
\sum_{i_{2k-1}}\Tr_{\spc H_{s_{2k-1}}}[R^{(k)}_{s_{2k-2}i_{2k-1}}]=I_{s_{2k-2}}\otimes R^{(k-1)}_{s_{2k-3}},
\end{equation}
where $R^{(N)}_{s_{2N-1}} :=R_{s_{2N-1}}$,
$R^{(k)}_{s_{2k-1}}\in\Lin{\bigotimes_{j=0}^{2k-1} \spc H_{s_j}} $,
and $R^{(0)} =1$.  Eq. (\ref{e:truegame}) is the mathematical
representation of the most general strategy in an $N$-round protocol
with exchange of classical and quantum information, generalizing the
game theoretical framework introduced by Gutoski and Watrous
\cite{watrous} for protocols involving only exchange of quantum
systems. We will call the collection of probabilistic quantum combs
satisfying Eq.  (\ref{e:truegame}) a {\em conditional comb}. This
nomenclature reflects the fact that the most general way of
conditioning a quantum comb needs to use at each step the information
coming from all previous steps.\par

Eq. (\ref{e:truegame}) sets a one-to-one correspondence between
single-party strategies in a protocol and conditional combs: indeed, a
collection of positive operators satisfying Eq.~\eqref{e:truegame} can
always be realized by a sequence of quantum instruments conditioned by
classical information, as in Fig. (\ref{f:ccomb}).  This fact is
proved in the following proposition.
\begin{Theo} {Any conditional comb is the collection of Choi-Jamio\l
    kowski operators of a sequential network of $N$ conditional
    instruments as in Fig. \ref{f:ccomb}.}
\end{Theo} 
\Proof Suppose that a collection of operators
$\{R_{s_{2N-1}}\}$ labeled by classical strings $s_{2N-1} = i_0 i_1 \dots i_{2N-1}$ satisfies conditions Eq.~\eqref{e:truegame}. Then,
we can define the operator
\begin{equation}
\begin{split}
  R:= \sum_{s_{2N-1}} & R_{s_{2N-1}}\otimes|s_{2N-1}\>\<s_{2N-1}| \\
 &\otimes|s_{2N-2}\>\<s_{2N-2}| \otimes \dots  \otimes |s_{0}\>\<s_{0}| .
\end{split}
\end{equation}
Here, $R$ acts on the tensor product $\spc K :=\bigotimes_{j=0}^{2N-1}{\spc H_j}$, where the $j$-th space is
$\spc H_j := \left(\bigoplus_{s_j}\spc H_{s_j}\otimes|s_j\>\right)$. With this definition, $R$ is a
deterministic comb, i.e. an operator satisfying Eq. (\ref{recnorm}). Therefore, by Proposition
\ref{prorea} $R$ can be realized with a network of $N$ channels $(\map C_k)_{k=0}^{N-1}$ as in Fig.
\ref{memch}. Now, if we apply the von Neumann-L\"uders measurements $ \{I_{s_{2k}} \otimes
|s_{2k}\>\< s_{2k}|\}$ on the input space $\spc H_{2k}$ before channel $\map C_k$, followed by $\{
I_{s_{2k+1}} \otimes |s_{2k+1}\>\< s_{2k+1}|\}$ on the output space $\spc H_{2k+1}$ after channel
$\map C_k$, we obtain the conditional quantum operations $\{\map C_{i_{2k+1}}^{s_{2k}}\}$.
Denoting by $C^{s_{2k}}_{i_{2k+1}}$ the Choi-Jamio\l kowski operator of the quantum operation $\map
C^{s_{2k}}_{i{2k+1}}$ we then have $R_{s_{2N-1}} = C^{s_{2N-2}}_{i_{2N-1}} * C^{s_{2N-4}}_{i_{2N-3}}
* \dots * C^{s_{0}}_{i_{1}} $, i.e. $R_{s_{2N-1}}$ is the Choi-Jamio\l kowski operator of the
sequence of quantum operations $(\map C^{s_{2k}}_{i_{2k+1}})_{k=0}^{N-1}$, as in Fig. \ref{f:ccomb}.
\qed

In the following we will consider the \emph{dilation of a conditional comb}
$\{R_{s_{2N-1}}\}$ defined as the collection $\{\widetilde
R_{s_{2N-1}}\}$ of dilations $\widetilde
R_{s_{2N-1}}\in\Lin{\left[\bigotimes_{j=0}^{2N-1} \spc
    H_{s_j}\right]\otimes\spc H_{A, s_{2N-1}}}$ of each comb
$R_{s_{2N-1}}\in\Lin{\bigotimes_{j=0}^{2N-1} \spc H_{s_j}}$, where
$\spc H_{A, s_{2N-1}}$ is an ancillary space depending on history. The
following theorem guarantees that the dilation of a conditional comb
is still a conditional comb.

\begin{Theo} {For any conditional comb $\{R_{s_{2N-1}}\}$ the dilation $\{\widetilde
    R_{s_{2N-1}}\}$ defined by $\widetilde  R_{s_{2N-1}} :=|R_{s_{2N-1}}^\frac12\kk\bb R_{s_{2N-1}}^\frac12|$ is a conditional comb.}
\end{Theo}

\Proof Define $\spc H'_{s_{2N-1}}:=\spc H_{s_{2N-1}}\otimes \spc H_{A, s_{2N-1}}$ and $\spc H_{s_l}' = \spc H_{s_l}$ for $l < 2N-1$. Then,
the operators $\{\widetilde R_{s_{2N-1}}\}$ form a conditional comb with  $\widetilde R_{s_{2N-1}} \in \Lin{\bigotimes_{j=0}^{2N-1}\spc
  H'_{s_j}}$. \qed

The dilation of a conditional comb describes as a sequence of
single-Kraus quantum operations, each of them depending on the
previously exchanged classical information.  Loosely speaking, this
theorem means that the ``quantum part'' of any strategy can be
purified until the end of the protocol, still resulting in a valid
strategy.

\section{Comb formulation of the quantum bit commitment}\label{s:mathform}  

A (generally multiparty) protocol establishes which are the honest single-party
strategies. A strategy is a choice of processing of
classical/quantum information at each step, and specifies which
quantum instrument a party will perform jointly on his ancillae and on
the received quantum systems, conditionally on the available classical
information.  
The honest strategies of the protocol fix the communication interface
among parties, consisting of the complete specification of which
classical and quantum systems are exchanged at each step.  A cheating
strategy can be any strategy that conforms to the communication
interface.

A definition of security of a protocol generally depends on the
specific goals of the involved parties. For the quantum bit commitment
a protocol is defined as perfectly secure if the following conditions
are satisfied:
\begin{itemize}
\item[] {\em concealment:} for all Alice's honest  strategies Bob
  cannot read the committed bit before the opening;
\item[] {\em bindingness:} for all honest Bob's strategies Alice cannot change the value of the
  committed bit without being detected. 
\end{itemize}
Note the asymmetry between the security condition for the two parties: on the one hand, security for Alice means that Bob
has no chance at all to read the bit, while, one the other hand, security for Bob means that if Alice tries to cheat, she will be surely detected. Perfect
security is relaxed to the case of {\em $\varepsilon$-concealment} and {\em $\delta$-bindingness}, where the
probability for Bob to read the committed bit is bounded by $\varepsilon$, and the
probability for Alice to change the bit value is bounded by $\delta$.

In the following subsections we will formulate strategies in terms of quantum combs, and evaluate
the probabilities of successfully cheating for both parties.
\subsection{Alice's and Bob's strategies}
As already noticed, there is no loss of generality in considering bit
commitment protocols started by Bob. With the letter $k=1,\ldots, N$
we will denote the $k$-th Bob's and Alice's step. Thus
$s_l=i_0i_1\ldots i_l$ will represent the history of classical
information with $i_{2k-1}$ denoting the outcome of Bob's quantum
operation at step $k$ (which is the same as Alice's classical input at
Alice's step $k$) and $i_{2k-2}$ for $k>1$ represents Bob's input
classical information (which is Alice's outcome at step $k-1$). At the
beginning of the protocol there is no classical and quantum
information, whence $s_0=i_0$ is the null string and $\spc H_0=\mathbb
C$. At the end of the commitment stage we can assume without loss of
generality that Alice performs the last move (for a protocol where the
last move is Bob's, we can always add a null move, in which no
classical and quantum systems are sent).

We now analyze the case in which the total number of steps in the
protocol is bounded uniformly over all possible histories, and
denote by $N$ the maximum number of steps.  Moreover, since we can
always add null moves, we  consider without loss of generality
protocols where the number of steps is $N$ independently of the
history.  Therefore the classical history labeling the sequence of
quantum operations will be $s_{2N}$ for Alice, and $s_{2N-1}$ for Bob.
Finally, by adding null steps we can decide without loss of generality
that in the last move before the opening Alice performs just a local
operation on her ancillae, \emph{i.e.} she does not send to Bob any
classical or quantum information. Accordingly, $i_{2N}$ is the null string,
and $\spc H_{s_{2N}} = \mathbb C$ for any history $s_{2N}$. Since both $i_0$ and $i_{2N}$ are null strings, we have $s_{2N} \equiv s_{2N-1} \equiv i_1 i_2 \dots i_{2N-1}$.

We denote by $\mathsf A_0$ and $\mathsf A_1$ the sets of honest
strategies that Alice can use to encode bit values 0 and 1,
respectively.  According to Subect.  \ref{s:ccombs}, a possible
strategy in $\mathsf A_i$ is a conditional quantum comb
$\{A_{i,s_{2N}}\}$, where the index $s_{2N}$ labels a history of
classical information exchanged between Alice and Bob. For each
history $s_{2N}$, $A_{i,s_{2N}}$ is a probabilistic comb on $\spc
K_{s_{2N}} \otimes \spc H_{A,s_{2N}}$, where $\spc
K_{s_{2N}}=\Lin{\bigotimes_{j=1}^{2N} \spc H_{s_j}} $ is the Hilbert
space of all quantum systems exchanged in the protocol and $\spc
H_{A,s_{2N}}$ is the Hilbert space of Alice's private ancillae at the
last step of the commitment phase.

\medskip We denote by $\mathsf B$ the set of strategies (honest or
not) that are available to Bob. The set $\mathsf B$ can be the whole
set of strategies compatible with the communication interface, or a
restricted subset. The only assumption here is that if $\mathsf B$
contains a strategy, then it contains also its dilations.  The reader
should then regard $\mathsf B$ as a parameter of his own choice for
the rest of the paper: the impossibility proof will state
that if the protocol is concealing for a Bob restricted to $\mathsf
B$, then it is necessarily not binding for Bob restricted to that
subset.   An element of
$\mathsf B$ is a collection of probabilistic quantum combs
$\{B_{s_{2N-1}}\}$.  For each history $s_{2N-1}$, $B_{s_{2N-1}}$ is a comb on
$\spc K_{s_{2N-1}} \otimes \spc H_{B,s_{2N-1}}$, where $\spc K_{S_{2N-1}} : = \bigotimes_{j=0}^{2N-1}  \spc H_{s_j}$, and $\spc H_{B,s_{2N-1}}$
is the Hilbert space of Bob's ancillae at the last step of the commitment phase. Note that, since $\spc H_{s_0} = \spc H_{s_{2N}} = \mathbb C$, one has $\spc K_{s_{2N}} \simeq \spc K_{S_{2N-1}} \simeq \bigotimes_{j=1}^{2N-1} \spc H_{s_j}$.

\medskip

\medskip 

In the following we focus on the last step $N$ before the opening.
Since the step is fixed, we will drop the sub-index $2N$ ($2N-1$)
labeling Alice's (Bob's) history.  For the history $s$, the overall
(unnormalized) state resulting from Alice and Bob playing the
strategies $\{A_{i,s'}\}$ and $\{B_{s'}\}$, respectively, is given by
the link product
\begin{equation}
\sigma^{(i)}_s  = B_s * A_{i,s} \in \Lin{\spc H_{A,s} \otimes \spc H_{B,s}}. 
\end{equation}  
The probability of the history $s$ is then given by the trace
\begin{equation}
p^{(i)}_{s} = \Tr [\sigma^{(i)}_s]
\end{equation}  
The local state at Bob before the opening is
\begin{equation}
\rho^{(i)}_s  = \Tr_{\spc H_{A,s}}  [\sigma^{(i)}_s] = B_s * R_{i,s} ,\end{equation}
where  
\begin{equation} R_{i,s} = \Tr_{\spc H_{A,s}} [A_{i,s}] \in\Lin{\spc K_s}.
\end{equation} 
is the restriction of Alice's comb to the quantum systems exchanged in
the protocol.

\subsection{Concealing protocols}


\begin{Def}[Concealing protocols]
  A quantum bit commitment protocol is $\varepsilon$-concealing if there is at least a couple of honest
  strategies $\{A_{0,s'}\} \in \mathsf A_0 $, $\{A_{1,s'}\} \in \mathsf A_1$ such that
  the following conditions hold:
\begin{eqnarray}
  \max_s \frac{\left \|  \rho^{(1)}_s - \rho^{(0)}_s \right\|_1}{\Tr\left[  \rho^{(1)}_s + \rho^{(0)}_s \right]} & \le&
  \varepsilon , \quad\forall \{B_{s'}\}\in\mathsf B 
\label{conce}
\end{eqnarray}
where $\rho^{(i)}_s$ is the unnormalized state on Bob's side $\rho^{(i)}_s =
B_s * R_{i,s}$, with $R_{i,s} = \Tr_{\spc H_{A,s}} [A_{i,s}]$.
\end{Def}

As discussed in subsection \ref{subsect:discriminability}, the the above condition means that, for any history of classical communication, the probability that Bob discriminates correctly between $R_{0,s}$ and $R_{1,s}$ is $\varepsilon$-close to $1/2$, the success probability of a random guess.    


\medskip
The concealment condition can be translated in terms of
combs distances as follows:
\begin{Lem}\label{lem:conc}  A protocol is $\varepsilon$-concealing if and only if there is a couple of
  honest strategies $\{A_{0,s'}\}$ and $\{A_{1,s'}\}$ such that
  \begin{equation}\label{conceal}
    \max_s  \left.d ( R_{1,s} , R_{0,s})\right|_{\mathsf T_s}  \le {\varepsilon} .
  \end{equation}
where ${\mathsf T_s}=\{T_s:= \Tr_{\spc H_{B,s}} [B_s],\, B_s\in\{B_{s'}\}\in\mathsf{B} \}$.
\end{Lem} 
\Proof Clearly, condition (\ref{conce}) holds if an only if
\begin{equation}
 \max_s \sup_{\mathsf B_s} \frac{\|  \rho^{(1)}_s -  \rho^{(0)}_s \|_1}{ \Tr\left[\rho^{(1)}_s + \rho^{(0)}_s\right]} \le \varepsilon,
\end{equation}  where ${\mathsf B_s}=\{B_s\in\{B_{s'}\} \in\mathsf{B} \}$. 
Moreover, since the set of Bob's strategies is closed under dilation, and since dilation improves the discrimination, the supremum can be taken over the dilations $\{\widetilde B_{s'}\}$. Now, denote by $\widetilde T_s$ the dilation of $T_s = \Tr_{\spc H_{B,s}} [B_s]$.
Since $\widetilde B_s$ and $\widetilde T_s$ are both dilations of $T_s$, they are connected by a
partial isometry on Bob's ancillae.  The same is true for the states $ \tilde \rho^{(i)}_s :=
\widetilde B_s * R_{i,s} $, and $\rho^{(i)}_{T_s} := \widetilde T_s *
R_{i,s}$, for each value $i = 0,1$, whence $\| \tilde \rho^{(0)}_s- \tilde \rho^{(1)}_s\|_1=
\| \rho^{(0)}_{T_s}- \rho^{(1)}_{T_s}\|_1$.  This implies the identity
\begin{equation}
\begin{split}
  \max_s \sup_{{\mathsf B_s}} \frac{\left\| \rho^{(1)}_s -  \rho^{(0)}_s
  \right\|_1 }{\Tr[\rho^{(1)}_s + \rho^{(0)}_s]} &= \max_s \sup_{ \mathsf B_s} \frac{\left\|    \tilde\rho^{(1)}_{s} -  \tilde \rho^{(0)}_{s} \right\|_1}{\Tr[\tilde \rho^{(1)}_{s} + \tilde \rho^{(0)}_{s}]}\\
&= \max_s \sup_{ \mathsf T_s} \frac{\left\|    \rho^{(1)}_{T_s} -  \rho^{(0)}_{T_s} \right\|_1}{\Tr[\rho^{(1)}_{T_s} + \rho^{(0)}_{T_s}]}\\
 & = \max_s\left. d (R_{1,s},R_{0,s}) \right|_{\mathsf T_s}.
\end{split}
\end{equation}
\qed

\subsection{Alice's cheating strategies}
Let $\{A_{s'}\}$ and $\{A_{s'}^\sharp\}$ be a honest and a dishonest
strategy by Alice, respectively (here we drop the index $i=0,1$ of the
bit value, since it is unnecessary for the following discussion).
When Bob chooses the strategy $\{B_{s'}\} \in \mathsf B$, for history
$s$ the unnormalized quantum states before the opening phase are
\begin{equation}
 \sigma_{s} = {B_s *
  A_s}, \qquad  \sigma^\sharp_{s} =
{B_s* A_{s}^\sharp}.
\end{equation}

\begin{Def} The strategy $\{A_{s'}^\sharp\}$ is $\delta$-close to the strategy $\{A_{s'}\}$ at the
  opening if for any strategy $\{B_{s'}\}\in \mathsf B$ one has
\begin{equation}\label{cheat}
\begin{split}
  \max_s \frac{\left\|  \sigma_{s} -  \sigma^\sharp_{s} \right\|_1}{\Tr\left[\sigma_s + \sigma_s^\sharp\right]}&
  \le \delta .
\end{split}
\end{equation}
\end{Def}
If two strategies are $\delta$-close, Bob cannot distinguish between
them,  even if the history that takes
place is the most favorable to him.

Following the same argument used in the proof of lemma \ref{lem:conc}, the notion of $\delta$-closeness can be expressed in terms of comb distance as follows: 
\begin{Lem}\label{betterdishonest}
  The strategy $\{A_{s'}^\sharp\}$ is $\delta$-close to the strategy
  $\{A_{s'}\}$ at the opening if and only if
\begin{equation}
\max_s  \left. d (A_s, A_s^\sharp) \right|_{\mathsf T_s \otimes I_{A,s}}\le \delta,
\end{equation}
where ${\mathsf T_s}\otimes I_{A,s}=\{T_s\otimes I_{A,s}| T_s=\Tr_{\spc
  H_{B,s}} [B_s],\, B_s\in\{B_{s'}\}\in\mathsf{B} \}$ and $I_{A,s}$
denotes the identity on Alice's ancilla $\spc H_{A,s}$.
\end{Lem}

\begin{Def}
Given two honest strategies $\{A_{0,s'}\} \in \mathsf A_0$ and  $\{A_{1,s'}\} \in \mathsf A_1$, a $\delta$-cheating is a couple of strategies $\{A^\sharp_{0,s'}\}$ and  $\{A^\sharp_{1,s'}\}$  satisfying the conditions
\begin{enumerate}
\item $\{A_{i, s'}^\sharp\}$ is $\delta$-close to  $\{A_{i,s'}\}$ for $i=0,1$
\item for every history $s$, there exists a quantum channel $\map C_s$ acting on Alice's ancilla space $\spc H_{A,s}$ such that
\begin{equation}
A_{1,s}^{\sharp} = (\map I_s\otimes \map C_s)(A_{0,s}^\sharp),
\end{equation}   
where $\map I_s$ is the identity channel on the Hilbert space $\spc
K_s$ of all quantum systems exchanged in the commitment phase.
\end{enumerate} 
\end{Def}
The second condition means that Alice can follow the strategy
$\{A^\sharp_{0,s'}\}$ until the end of the commitment, and switch to
the strategy $\{A_{1,s}^\sharp\}$ with a local operation on her
ancillae just before  the opening.
\section{The impossibility proof}

\subsection{Protocols with bounded number of rounds}

\begin{Theo}\label{theo:impossibility}
  If an $N$-round protocol  is $\varepsilon$-concealing with honest strategies
  $\{A_{0,s}\} \in \mathsf A_0$ and $\{A_{1,s}\} \in \mathsf A_1$,
  then there is a $\sqrt {2\varepsilon}$-cheating with cheating
  strategies $\{A^\sharp_{0,s}\} $ and $\{A^\sharp_{1,s}\}$. In particular, the cheating strategy $\{A_{0,s}^\sharp\}$ coincides with the honest
  strategy $\{A_{0,s}\}$.
\end{Theo}

\Proof According to Eq. (\ref{conceal}), the concealing condition is for any history $s$
\begin{equation}
\left.d(R_{1,s}, R_{0,s})\right|_{\mathsf T_s} < \varepsilon ,
\end{equation}
where $R_{i,s} = \Tr_{\spc H_{A,s}} [A_{i,s}]$ and $\mathsf T_s = \{
T_s = \Tr_{\spc H_{B,s}} [ B_s] ~|~ B_s \in\{B_{s'}\}\in \mathsf B\}$.
We now focus on a fixed history $s$, and show the existence of two
$\sqrt{2\varepsilon}$-cheating strategies $\{A_{0,s'}^\sharp\}$ and
$\{A_{1,s'}^\sharp\}$. Since we are fixing $s$, we drop the index $s$
everywhere.

Since the reduced combs $ R_i = \Tr_{\spc H_{A}} [A_{i}] \in \Lin{\spc
  K}$ satisfy the condition $\left. d(R_1,R_0)\right|_{\mathsf T} <
\varepsilon$, we can use the continuity of  dilation stated by Lemma
\ref{conticondcomb}, thus finding a random unitary channel $\mathcal
P=\sum_k p_k\mathcal U_k$ acting on the ancilla space $\spc H_A \simeq \spc K$ such that
\begin{equation}\label{contis}
  \left. d( \widetilde R_{1} , (\mathcal I\otimes\mathcal P)  \widetilde R_{0} )\right|_{\mathsf T\otimes I_A}
  \le   \sqrt{2\left. d ( R_{1} , R_{0} )\right|_{\mathsf T}}, 
\end{equation}   
where $\widetilde R_{i}$ is the dilation $\widetilde R_{i} = |R^{\frac 1 2}_{i} \kk \bb R^{\frac 1
  2}_{i}| \in \Lin {\spc K\otimes\spc K_A}$. Now consider the dilations
of the honest strategies
\begin{equation}
  \widetilde A_i := |A_{i}^{\frac 1 2} \kk \bb A_{i}^{\frac 1 2}|.  
\end{equation}
Here $\widetilde A_{i}$ is an operator in $\Lin{ \spc K\otimes \spc
  H_{A} \otimes \spc L_{A}}$ where $\spc L_{A} \simeq \spc K\otimes
\spc H_{A}$ is an additional ancilla space on Alice's side. By
definition, $R_{i} = \Tr_{\spc H_{A},\spc L_{A}} [\widetilde
A_{i}]$.  Since $\widetilde A_{i}$ and $\widetilde R_{i}$ are
both dilations of $ R_{i}$, there exist a channel $\mathcal
E_{i}$ sending states on $(\spc H_{A} \otimes \spc L_{A})$ to
states on $\spc K_A$ such that
\begin{equation}
  \widetilde R_{i} =  (\mathcal I_{\spc K} \otimes \mathcal E_{i}) (\widetilde A_{i}),
\end{equation} 
and a channel $\mathcal F_{i}$ sending states on $\spc K_A$ to states on $(\spc H_{A} \otimes \spc
L_{A})$ such that
\begin{equation}
  \widetilde A_{i} = (\mathcal I_{\spc K} \otimes \mathcal F_{i}) ( \widetilde R_{i}).
\end{equation}
Alice's cheating procedure is then the following: 
\begin{itemize}
\item Use the dilated strategy  $\widetilde A_0$
\item After the commitment decide the bit
  value. To commit $0$, do nothing. To commit $1$, apply the
  channel $ \map C = \mathcal F_{1} \mathcal P \mathcal E_{0}$ on the ancillae,
  where $\mathcal P (\rho) = \sum_ip_iU_i \rho U_i^\dag$.
\item Discard the additional ancilla $\spc L_{A}$.
\end{itemize}
This procedure defines for every history $s$  the two cheating strategies $\{A_{0,s'}^\sharp\} := \{A_{0,s'}\}$  and $\{A_{1,s'}\} := \{(\map I_{s'} \otimes \map C_{s'} )(A_{0,s'}^\sharp)\}$.  Clearly,  $\{A_{0,s'}^\sharp\}$ is $\sqrt {2\varepsilon}$-close to  $\{A_{0,s'}\}$  (in fact, they coincide). Regarding $\{A_{1,s'}^\sharp\}$, for any history $s$ (and hence dropping the index) we have 
\begin{equation}
\begin{split}
  \left. d( A_1, A_1^\sharp )\right|_{\mathsf T\otimes I_A}   &=\left. d\left(  A_{1} , \Tr_{\spc L_{A}} \left [ ( \mathcal I\otimes \mathcal F_{1} \mathcal P\mathcal E_{0} )  (\widetilde A_0)  \right]  \right)\right|_{\mathsf T\otimes I_A}\\
  &\le \left. d \left( \widetilde A_{1} ,  ( \mathcal I \otimes \mathcal F_{1} \mathcal P \mathcal E_{0} )  (\widetilde A_0) \right)\right|_{\mathsf T\otimes I_A }\\
  &= \left. d\left(  (\mathcal I\otimes \mathcal F_{1}) (\widetilde R_1) ,  (\mathcal I \otimes \mathcal F_1 \mathcal P) (\widetilde R_{0}) \right)  \right|_{\mathsf T\otimes I_A}\\
  &\le \left.  d \left( \widetilde R_{1} ,(\mathcal I\otimes \mathcal
      P) (\widetilde R_{0}) \right) \right|_{\mathsf
    T\otimes I_A}\\
  & \le \sqrt{2\left. d(R_1, R_0)\right|_{\mathsf T}}\le \sqrt{
    2\varepsilon}.
\end{split}
\end{equation}
 Here, the first and the second inequalities derive from Lemma \ref{l:trc-dist},  the third one is
 Eq. (\ref{contis}), and the last is  the concealing condition. \qed 

\subsection{Protocols with unbounded number or rounds} 
Here we show how the impossibility result of the previous subsection
can be easily extended to the case of protocols where the number of
rounds is unbounded.  In this case Alice's (Bob's) strategies are
still described by collections of probabilistic combs $\{A_{s'}\}$ and
($\{B_s'\}$), where each probabilistic comb represents the sequence of
quantum operations performed by Alice (Bob) for a given history $s$ of
classical communication.  Note that, although the length the strings
is no longer bounded by a fixed number, any given string
$s$ must have finite length. Indeed, a protocol allowing an infinitely
long history $s$ would be a protocol in which sometimes Alice and Bob
have to continue their communication forever, without reaching neither
a successful commitment, nor an abort.

For a protocol with unbounded number of rounds, the conditions of $\varepsilon$-concealment and
$\delta$-closeness are still given by Eqs. (\ref{conce}) and (\ref{cheat}), respectively.  Now, it
is immediate to see that, given an $\varepsilon$-concealing protocol with unbounded number of rounds,
one can always construct a new $\varepsilon$-concealing protocol with bounded number. Indeed, Alice
can follow the original unbounded protocol, and decide to abort whenever the number of rounds
exceeds a fixed number $N$.  This change does not change the security of the protocol: it just
reduces the probability of successful commitment by turning some histories that in the original
protocol ended in a successful commitment into histories that end in an abort.  For the new protocol
with finite rounds, however, one can apply theorem \ref{theo:impossibility}, thus finding a
$\sqrt{2\varepsilon}$-cheating for Alice. Since $N$ is arbitrary and since for any $N$ the cheating
strategy coincides with the honest one up to the opening, Alice can take the number $N$ to be
sufficiently large to make the probability of successful commitment close to the one of the unbounded original
protocol.


\section{Summary}\label{s:conc}

In this paper we have provided a new short impossibility proof of
quantum bit commitment. The present proof differs from the previous
ones in the following main aspects: {\em a)} The strategies, including
all their ``purifications'', have a simple and univocal mathematical
representation in terms of conditional quantum combs in Eq.
(\ref{e:truegame}); {\em b)} The definition of concealment and
bindingness are worst-case over histories, namely the conditions on
cheating probabilities are defined uniformly over histories of
classical communication rather than on average; {\em c)} we consider
the possibility of restricting the strategies of Bob to an arbitrary set closed under dilation, and show that if the protocol is
concealing for Bob restricted in this way, then it is not binding.
Along similar lines it is possible to prove the impossibility theorem
also with cheating probabilities averaged over histories.  However,
the two impossibility theorems are not comparable, since worst-case
concealment implies concealment in average, whereas bindingness in
average implies worst-case bindingness.

At the end of the paper, we want to stress two points regarding
abortion probabilities. First, concealment is defined regardless
abortion, namely Bob must not be able to detect the bit value anyway,
whether Alice catches him or not. Second, in order to cheat Alice has
only to play the honest strategy $\{A_{0,s'}\}$ up to the very last
moment before the opening, at which point her cheat is anyway
undetectable by Bob (at the opening Bob's success probability in
detecting the cheat is at most $\sqrt{2 \varepsilon}$-close to the
success probability of a random guess).  Therefore, the probability of
abort before the opening is independent on whether Alice is cheating
or not.



\begin{acknowledgments}
  GMD acknowledges interesting discussions with H. Yuen, H. K. Lo, and C. Bennett. GC, GMD, PP
  acknowledge A. Bisio for useful observations.
This work has been supported by EC through projects CORNER and COQUIT.
\end{acknowledgments}

\appendix

\section{Proof of minimax equality in Eq. (\ref{minimaxbound})}\label{app} 
\begin{Lem}
Let $f$ be the function from $\mathsf C \times \mathsf T$ to $\mathbb R$: 
\begin{align}
f(C,T) &= \mathrm{Re} \bb \Psi^{(1)}_{T,I} | \Psi^{(0)}_{T,C} \kk\\
 \bb \Psi^{(1)}_{T,I} | \Psi^{(0)}_{T,C} \kk &= \frac{  \bb R^{\frac 1 2}_1 | T^\tau \otimes C | R^{\frac 1 2}_0 \kk} {\Tr[T^\tau (R_0+R_1)]}
\end{align}
where $R_i \ge 0$.  Then one has the identity
\begin{equation}
{\inf_{T\in \mathsf T}}'\sup_{C\in \mathsf C}  f(C,T) =\sup_{C\in \mathsf C} {\inf_{T\in \mathsf T}}'  f(C,T), 
\end{equation} 
where the infimum over $T$ is taken over the set of tester operators $T \in \mathsf T$ such that $\Tr [T^\tau (R_0+R_1)]\not = 0$.
\end{Lem}
\Proof Let $\overline{\mathsf T}$ be the closure of the convex hull of
$\mathsf T$. Since we are in finite dimensions, $\overline {\mathsf
  T}$ is a compact set. Define the compact convex set $\mathsf T_n $
as
\begin{equation}
  \mathsf T_n :=\left\{  T \in \overline{\mathsf T} ~|~ \Tr [T^\tau (R_0+R_1)] \ge \frac 1 n \right\} . 
\end{equation}
We now restrict $f$ to the set $\mathsf T_n$ and apply Sion's minimax
theorem \cite{fan}. The hypotheses of the theorem are satisfied:
First, the function is continuous versus $C$ and $T$, and both sets
$\mathsf C$ and $\mathsf T_n$ are compact and convex.  Finally, the
function $f$, being linear-fractional, is {\em quasi-linear} in $T$
for every $C$ \cite{boyd}, whereas it is linear in $C$ for every $T$
in its domain.  For arbitrary $n$, Sion's theorem \cite{fan} yields
the equality
\begin{equation}
\begin{split}
\inf_{T \in \mathsf T_n} \sup_{C\in \mathsf C}  f(C,T)=\sup_{C\in \mathsf C} \inf_{T \in \mathsf T_n}  f(C,T) 
\end{split}
\end{equation}
Moreover, the function $g(T) := \sup_{C \in \mathsf C} f(C, T)$ is
quasi-concave, since it is the supremum over one variable of a jointly
quasi-concave function of $(C,T)$ \cite{boyd}.  Therefore, the infimum of $g$ over $\mathsf
T$ is equal to the infimum over the closed convex hull
$\overline{\mathsf T}$, and we have
\begin{equation}
\begin{split}
  &{\inf_{T \in \mathsf T}}'\sup_{C\in \mathsf C}  f(C,T)= {\inf_{T \in \mathsf T}}'  g(T) =    {\inf_{T \in \overline{\mathsf T}}}'  g(T) \\
  &= {\inf_{T \in \overline{\mathsf T}}}'\sup_{C\in \mathsf C} f(C,T) = \inf_n\inf_{T \in \mathsf T_n}\sup_{C\in \mathsf C} f(C,T)\\
  &= \inf_n\sup_{C\in \mathsf C} \inf_{T \in \mathsf T_n} f(C,T) \ge
  \sup_{C\in \mathsf C} {\inf_{T \in \overline{ \mathsf T}}}' f(C,T) \\
& =  \sup_{C\in \mathsf C} {\inf_{T \in \mathsf T}}' f(C,T) ,
\end{split}
\end{equation} 
having used that $f(C,T)$ is quasi-concave in $T$ for any $C$, whence the infinum ove $\overline{\mathsf T}$ is equal to the infimum over $\mathsf T$. 
In fact, the above bound is achieved: For any $\varepsilon
>0$, and for any $C \in \mathsf C$, there is an element $T_C \in
\mathsf T$ such that $f(C, T_C) \le {\inf'_{T \in \mathsf T}} f(C,T) +
\varepsilon/2$.  Moreover, since $f$ is continuous in its domain, there exist two open sets  $A_C\subseteq \mathsf C$ and  $ B_{T_C}\subseteq \mathsf T$  such that for $C'\in A_C$ and $T'\in B_{T_C} $ one has $
f(C', T') \le f(C, T_C) + \varepsilon/2$.
Now, the open sets $\{A_C\}$ form an open cover of $\mathsf C$.
Since $\mathsf C$ is compact, one can extract from $\{A_C\}$
a finite subcover $\{A_{C_i}\}$.  Finally, for any $i$ there is a
number $n_i$ such that the intersection between $B_{T_{C_i}}$ and $
\mathsf T_{n_i}$ is non empty. Let us define $n_{\varepsilon} =
\max\{n_i\}$. Then we have
\begin{equation}
\begin{split}
&\sup_{C\in A_{C_i}}{\inf_{T \in T_{n_{\varepsilon}}}} f(C, T)  \le \sup_{C\in A_{C_i}}\inf_{T \in T_{n_\varepsilon} \cap B_{T_i}} f(C, T)\\
&  \le f(C_i, T_{C_i}) + \varepsilon/2
\le {\inf_{T \in \mathsf T}}' f(C_i, T) + \varepsilon \\ & \le \sup_{C} {\inf_{T \in \mathsf T}}' f(C, T) + \varepsilon.  
\end{split}
\end{equation}    
Since the sets $\{A_{C_i}\}$ cover $\mathsf C$, this implies $\sup_{C\in \mathsf C} \inf_ {T_{n_\varepsilon}}  (f(C,T))\le \sup_{C \in \mathsf C} {\inf'_{T \in \mathsf T}} f(C, T)  + \varepsilon$, whence
\begin{equation}
\begin{split}
&{\inf_{T \in \mathsf T}}' \sup_{C \in \mathsf C} f(C,  T) = \inf_n \sup_{C \in \mathsf C} \inf_{T \in \mathsf T_n} f(C, T) \\ &\le\sup_{C \in \mathsf C} {\inf_{T \in \mathsf T}}' f(C, T) + \varepsilon.  
\end{split}
\end{equation} 
\qed 


\bigskip

\begin{thebibliography}{99}
\bibitem{LC97} H. K. Lo, H. F. Chau, Phys. Rev. Lett. {\bf 78}, 3410
  (1997).
\bibitem{May97} D. Mayers, Phys. Rev. Lett. {\bf 78}, 3414 (1997).
\bibitem{werqbc} G. M. D'Ariano, D. Kretschmann, D. Schlingemann, and R.
  F. Werner, Phys. Rev. A {\bf 76}, 032328 (2007).
\bibitem{qca} G. Chiribella, G. M. D'Ariano, and P. Perinotti, Phys. Rev. Lett. {\bf 101}, 060401 (2008).
\bibitem{Kil88} J. Kilian, in {\em Proceedings of the 20th ACM
    Symposium on Theory of Computing}, (ACM, New York, 1988), p. 20.
\bibitem{Blu83} M. Blum, SIGACT News {\bf 15}, 23 (1983).
\bibitem{BBC+91} C. H. Bennett, G. Brassard, C. Cr\'epeau, M. H.
  Skubiszewska, in {\em Advances in Cryptology --- Proceedings of
  CRYPTO'91}, (Springer, Berlin, 1991), p. 351.
\bibitem{Cre94} C. Cr\'epeau, J. Mod. Opt. {\bf 41} 2455 (1994)
\bibitem{Yao95} A. C. C. Yao, in {\em Proceedings of the 27th ACM
    Symposium on Theory of Computing}, (ACM, New York, 1995), p. 67.
\bibitem{CVT95} C. Cr\'epeau, J. van de Graaf, A. Tapp, in {\em
    Proceedings of the 15th Annual International Cryptology Conference
    on Advances in Cryptology (CRYPTO'95)}, Lect. Notes in Computer
  Science {\bf 963}, (Springer, Berlin, 1995), p. 110.
\bibitem{BB84} C. H. Bennett, G. Brassard, in {\em Proceedings of IEEE
    International Conference on Computers, Systems, and Signal
    Processing, Bangalore, India, 1984}, (IEEE, New York, 1984), pp.
  175-179.
\bibitem{Eke91} A. K. Ekert, Phys. Rev. Lett. {\bf 67}, 661 (1991).
\bibitem{BCJ+93} G. Brassard, C. Cr\'epeau, R. Jozsa, D. Langlois, in
  {\em Proceedings of the 34th Annual IEEE Symposium on the
    Foundations of Computer Science}, (IEEE Computer Society Press,
  Los Alamitos, 1993), p. 362.
\bibitem{Uhl76} A. Uhlmann, Rep. Math. Phys. {\bf 9}, 273 (1976).
\bibitem{KMP04} A. Kitaev, D. Mayers, J. Preskill, Phys. Rev. A {\bf
    69}, 052326 (2004).
\bibitem{Yuenall} H. P. Yuen, quant-ph/0006109, 0305144, 0505132,
  0702074
\bibitem{Yuenlast} H. P. Yuen, arXiv:0808.2040.
\bibitem{Che01} C. Y. Cheung, quant-ph/0112120.
\bibitem{Kraus} K. Kraus, Ann. Phys. {\bf 64}, 311 (1971).
\bibitem{comblong} G. Chiribella, G. M. D'Ariano, and P. Perinotti,  arXiv:0904.4483.
\bibitem{watrous} G. Gutoski and J. Watrous, in {\em Proceedings of
    the Thirtyninth Annual ACM Symposium on Theory of Computation
    (STOC)}, pag. 565-574 (2007).
\bibitem{memorydisc} G. Chiribella, G. M. D'Ariano, and P. Perinotti, Phys. Rev. Lett. {\bf 101}, 180501 (2008).
\bibitem{Ken99} A. Kent, Phys. Rev. Lett. {\bf 83}, 1447 (1999).
\bibitem{Ken05} A. Kent, J. Cryptology {\bf 18}, 313 (2005).
\bibitem{Stinespring}
W.~F. Stinespring.
\newblock {\em Proc. Amer. Math. Soc.}, 6:211--216, 1955.
\bibitem{Ozawa} M.~Ozawa.
\newblock {\em J. Math. Phys.}, 25:79, 1984.
\bibitem{Kretschmann:2008p3022} D. Kretschmann, D. Schlingemann, and R. F. Werner, J. Funct. Anal.
{\bf 255}, 1889 (2008)
\bibitem{fan} M. Sion, Pac. J. Math. 8 171 (1958)
\bibitem{boyd} S. P. Boyd and L. Vandenberghe, {\em Convex Optimization}, (Cambridge University
  Press, Cambridge 2008)
\end{thebibliography}
\end{document}